\documentclass[aps,prb,twocolumn,showpacs,amsmath]{revtex4}
\usepackage{latexsym}
\usepackage{amssymb}
\usepackage{graphicx}

\newcommand{\pdag}{{\phantom{\dagger}}}
\newcommand{\bq}{\begin{equation}}
\newcommand{\eq}{\end{equation}}
\newcommand{\bn}{\begin{eqnarray}}
\newcommand{\en}{\end{eqnarray}}

\begin{document}

\title{Quantum rate equations for electron transport through an interacting system in 
the sequential tunneling regime}

\author{Bing Dong} 
\affiliation{Department of Physics and Engineering Physics, Stevens Institute of 
Technology, Hoboken, New Jersey 07030 \\
Department of Physics, Shanghai Jiaotong University, 1954 Huashan Road, Shanghai 200030, 
China}
\author{H. L. Cui} 
\affiliation{Department of Physics and Engineering Physics, Stevens Institute of 
Technology, Hoboken, New Jersey 07030}
\author{X. L. Lei} 
\affiliation{Department of Physics, Shanghai Jiaotong University, 1954 Huashan Road, 
Shanghai 200030, China}

\begin{abstract}

We present a set of modified quantum rate equations, with the help of the nonequilibrium 
Green's function and slave-particle techniques along with the correct quantization, for 
description of the quantum transport through an interacting mesoscopic region connected 
with two leads, in the sequential tunneling regime. The assumption that only leading 
order of $|V|^2$ ($V$ is the tunneling coupling between the interacting central region 
and the leads) has been taken into account in deriving these equations implies that the 
quantum rate equations are only valid in the case of weak coupling between the central 
region and the leads. For demonstrations, we consider two special cases in the central 
region, a single interacting quantum dot (SQD) with weak spin-flip scattering and a 
weakly coupled double quantum dots (CQD), as examples. In the limit of zero temperature 
and large bias voltage, the resulting equations are identical to the previous results 
derived from the many-body Schr\"odinger equation. The numerical simulations reveal: 1) 
the dependence of the spin-flip scattering on the temperature and bias voltage in the 
SQD; and 2) the possible negative differential conductance and negative tunnel 
magnetoresistance in the CQD, depending on the hopping between the two quantum dots.

\end{abstract}


\maketitle

\section{Introduction}

For many years, much experimental and theoretical work has been devoted to exploring the 
transport properties of artificially nanofabricated structures containing a discrete 
number of quantum states and a small number of electrons. The tunneling current through 
these mesoscopic devices, isolated from two macroscopic leads by potential barriers, 
manifested many novel effects due to this confinement. For example, in a semiconductor 
quantum dot (QD) one observed the Coulomb blockade oscillations due to the charging 
energy,\cite{Averin} and even the Kondo effect due to the strong on-site Coulomb 
interaction in the tunneling transport.\cite{Ng,Kouwenhoven} Recently, interest in 
quantum computation and quantum information processing has attracted increasing 
attention to the problem of measurement of tunneling currents via a mesoscopic system 
that can be modeled by a two-level Hamiltonian, for example, charges in coupled 
QDs\cite{Vaart,Blick,Oosterkamp,Wiel} and spins in a QD under magnetic 
fields.\cite{Prinz} Measurement of the tunneling current in such systems provides 
information not only about the Rabi oscillations\cite{Wiel} between the two levels, but 
also about the spin precession in quantum spin oscillations,\cite{Manassen,Balatsky} 
both of which are crucial improvement in the control of the superposition of the quantum 
states and thus quantum information processing. In addition, similar physical picture 
has been utilized with success to analyze transport through molecular 
nanojuction,\cite{Hettler} for example, a system of benzene\cite{benzene} and DNA 
molecular chain.\cite{DNA}

In order to describe this kind of quantum oscillations in quantum transport through a 
QD, the master equations or ``quantum" version of the rate equations have been first 
proposed by Nazarov,\cite{Nazarov} and later derived microscopically from the 
Schr\"odinger equation directly,\cite{Gurvitz,Gurvitz1} and from the von Neumann 
equation and superoperators\cite{Engel}, respectively. In these, the central point is 
that the equations of motion (EOM) of the diagonal density matrix elements allow for an 
additional term of the nondiagonal density matrix elements, which indeed stand for the 
coherent superposition of different quantum states and are referred to as the coherent 
transfer term, along with the time evolutions commanded by their own EOM. We have to 
solve these equations selfconsistently to determine the nonequilibrium probability 
densities. As a result, the tunneling current unavoidably contains the contributions of 
the nondiagonal density matrix elements and naturally provides the information of the 
quantum Rabi oscillation, although the explicit expression of current formulation only 
involves the diagonal density matrix elements. The modified quantum rate equations have 
proved successful in describing this quantum oscillation in coherently coupled quantum 
dots (CQD),\cite{Nazarov} the quantum measurement by using quantum point contact near 
CQD,\cite{Korotkov,Goan,Elattari} and even time-dependent quantum tunneling through the 
CQD.\cite{Stoof} On the other hand, the Coulomb interaction inside the small confined 
region plays crucial role in, as above mentioned, determining the quantum transport 
properties of the devices, of course in controlling the quantum oscillations of 
two-level systems.\cite{Stafford} In fact, the so-called noninvasive quantum measurement 
process is also based on the Coulomb coupling between the detector and the measured 
system.\cite{Gurvitz2,Buttiker} To our knowledge, however, a systematic investigation of 
the quantum rate equations at arbitrary temperature associated with the Coulomb 
interactions has been lacking.

About ten years ago, a ``classical" rate equation was derived for sequential tunneling 
through a double-barrier system from the ``quantum" kinetic equation, the nonequilibrium 
Green's function (NGF), which is believed to be a more powerful tool for studying 
nonequilibrium phenomena.\cite{Davies,Hyldgaard,Hershfield} Our aim in this paper is to 
systematically explore the quantum rate equations for the interacting systems in the 
sequential tunneling regime from the NGF approach. The unique source of difficulty is 
how to deal with the Coulomb interaction term in the derivation. This problem is the 
same as that we have in studies of the strongly correlated fermionic systems, for 
example, the recent investigations on the Kondo enhanced conductance of a QD at low 
temperature.\cite{Ng,Hershfield,Wingreen,Aguado,Dong} Many theoretical methods have been 
developed to solve the strong correlation effects. Among these methods, the 
slave-particle technique is of particular elegance.\cite{Wingreen,Aguado,Dong,Hewson} 
The great advantage of this approach is that the correlated Hamiltonian for the system 
under studied is transformed to an equivalent one without Coulomb correlations, while 
introducing several auxiliary particles. Thus previously well-developed formulations for 
noninteracting systems can be applied to investigate the interacting systems in the 
framework of the new representation. Along this line, a further technique advance is 
made in the present work. Here we extend the approach of the slave-particle 
representation to the weakly coupled quantum system of interest here and give the 
consistent Hamiltonian formulation in terms of the slave particles. The equations of 
motion of the density matrix elements are then studied in the framework of NGF and 
within the slave-particle scheme. Our derivation contains three approximations. The 
first is to assume that the central region has very ``weak" coupling with the external 
environments (the leads) $V$. Secondly, we assume the couplings between subsystems are 
also weak to keep them being individual, for example, the weak spin-flip scattering in 
SQD and weak interdot hopping in CQD. As a result, we can give the definitions of the 
spectrum expressions of the NGFs of the central region in terms of the nonequilibrium 
probability densities and keep only the leading order term in $|V|^2$ in the expansions 
of the equations of motion. The final one is to apply the wide band limit in the two 
leads, namely that the coupling strengths between the central region and the leads are 
independent of the energy and can be considered as constant. 

The rest of the paper is organized as follows. In the next section, we give the 
derivations in detail for a single interacting QD (SQD) taking the weak spin-flip 
scattering into account, and establish the temperature and bias voltage dependent 
quantum rate equations for arbitrary Coulomb interaction. In section III we derive the 
quantum rate equations for the weakly coupled QDs. In both of the two section, after the 
analytical results are discussed for the no doubly occupied level and the deep level 
situations, we perform numerical simulations on the occupation numbers and the tunneling 
current in the general case as functions of the bare level in the QD and the bias 
voltage between the source and the drain. Finally, all the results are summarized in 
Section V.                      

\section{Single quantum dot}

We begin with our derivation of the quantum rate equations for a SQD with a weak 
spin-flip scattering in this section. In the case of no spin-flip terms, the rate 
equations are of ``classical" variety, which have been adequately described by other 
methods. Therefore our results are not new in this case but are established from a 
different scheme. The purpose of this section is also to provide an examination to prove 
this approach in comparison with previous results in no spin-flip case.   

\subsection{Model Hamiltonian and slave-particle representation}

We use the standard model Hamiltonian to describe the resonant tunneling through a SQD, 
as depicted in Fig.~1(a), with a single bare-level $\epsilon_{d}$ and a weak intradot 
spin-flip scattering $R_{\sigma}$ connected to two non-interacting leads:
\bn
H &=& \sum_{\eta, k, \sigma}\epsilon _{\eta k\sigma} 
c_{\eta k\sigma }^{\dagger }c_{\eta k\sigma }^{\pdag}+ \epsilon_{d} \sum_{\sigma} 
c_{d \sigma }^{\dagger }c_{d \sigma }^{\pdag}+ R_{\uparrow} c_{d \uparrow}^{\dagger} 
c_{d\downarrow}^{\pdag} \cr 
&& + R_{\downarrow} c_{d\downarrow}^{\dagger} c_{d\uparrow}^{\pdag} + Un_{d \uparrow 
}n_{d \downarrow }+\sum_{\eta, k, \sigma} (V_{\eta \sigma} c_{\eta k\sigma }^{\dagger 
}c_{d \sigma}^{\pdag} +{\rm {H.c.}}), \cr
&&
\label{hamiltonian1}
\en
where $c_{\eta k \sigma}^{\dagger}$ ($c_{\eta k \sigma }$) and $c_{d \sigma}^{\dagger}$ 
($c_{d \sigma}$) are the creation (annihilation) operators for electrons with momentum 
$k$, spin-$\sigma$ and energy $\epsilon_{\eta k \sigma}$ in the lead $\eta$ ($={\rm 
L,R}$) and for a spin-$\sigma$ electron on the QD, respectively. The third term 
describes the Coulomb interaction among electrons on the QD. $n_{d\sigma}=c_{d 
\sigma}^{\dagger} c_{d \sigma}^{\pdag}$ is the occupation operator in the SQD. The 
fourth term represents the tunneling coupling between the QD and the reservoirs. We 
assume that the coupling strength $V_{\eta \sigma}$ is spin-dependent, being able to 
describe the ferromagnetic leads. Each of the two leads is separately in thermal 
equilibrium with the chemical potential $\mu_{\eta}$, which is assumed to be zero in 
equilibrium condition and chosen as the energy reference throughout the paper. In the 
nonequilibrium case, the chemical potentials of the leads differ by the applied bias. In 
this paper, we assume the tunneling coupling is weak enough to guarantee no Kondo effect 
in our model and the QD is in the Coulomb blockade regime. Generally, we have 
$R_{\uparrow}=R_{\downarrow}^{*}=R$ being a constant.

According to the finite-$U$ slave-boson approach introduced by Zou and 
Anderson,\cite{Zou} the electron operator $c_{d \sigma}$ can be written in four possible 
single electron states, namely: the empty state $|0\rangle$ with zero energy 
$\varepsilon_{0}=0$, the singly occupied (with spin up or down) electronic state 
$|\sigma \rangle$ with energy $\varepsilon_{\sigma}=\epsilon_{d}$, and the doubly 
occupied state $| \uparrow\downarrow \rangle$ with energy 
$\varepsilon_{d}=2\epsilon_{d}+U$, as
\bq
c_{d \sigma}=|0\rangle \langle \sigma | + \sigma |\bar{\sigma} \rangle \langle \uparrow 
\downarrow |, \qquad (\sigma =\pm 1). \label{sf}
\eq
Because these four states expand the entire Hilbert space, the completeness relation 
must be satisfied
\bq
|0\rangle \langle 0|+ |\uparrow \downarrow\rangle \langle \uparrow \downarrow | + 
\sum_{\sigma} |\sigma \rangle \langle \sigma |=1. \label{comp}
\eq
These Dirac brackets were then treated as operators: $e^{\dagger}=|0\rangle$, 
$d^{\dagger}=|\uparrow \downarrow \rangle$ as slave-boson operators and 
$f_{\sigma}^{\dagger}=|\sigma \rangle$ as pseudo-fermion operator. In terms of these 
auxiliary operators, Eqs.(\ref{sf}) and (\ref{comp}) become
\bn
&&c_{d\sigma}=e^{\dagger} f_{\sigma}^{\pdag} + \sigma f_{\bar{\sigma}}^{\dagger} d, 
\label{sf2} \\
&&e^\dagger e + d^\dagger d + \sum_{\sigma}f_\sigma^\dagger f_\sigma =1.
\label{comp2}
\en
The explicit (anti)communicators of these auxiliary particles can be easily established 
from the definitions of the Dirac brackets:\cite{Guillou}
\bn
&& ee^{\dagger}=1, \quad dd^{\dagger}=1, \quad f_{\sigma}^{\pdag} 
f_{\sigma'}^{\dagger}=\delta_{\sigma \sigma'}, \cr
&& ed^{\dagger}=ef_{\sigma}^{\dagger}=f_{\sigma}^{\pdag}e^{\dagger} = f_{\sigma}^{\pdag} 
d^{\dagger} = de^{\dagger}=d f_{\sigma}^{\dagger}=0. \label{quan}
\en 
Therefore, along with these correct quantization, the Hamiltonian (\ref{hamiltonian1}) 
can be replaced by the following form in the auxiliary particle representation:
\bn
H&=& \sum_{\eta, k, \sigma }\epsilon _{\eta k \sigma } c_{\eta k \sigma }^{\dagger } 
c_{\eta k \sigma }^{\pdag} + \epsilon_{d} \sum_{\sigma} f_{\sigma }^{\dagger } f_{\sigma 
} \cr 
&& + (2\epsilon_{d}+U )d^\dagger d
+ R_{\uparrow} f_{\uparrow}^{\dagger} f_{\downarrow}^{\pdag}+ R_{\downarrow} 
f_{\downarrow}^{\dagger} f_{\uparrow}^{\pdag} \cr
&& + \sum_{\eta, k, \sigma} [ V_{\eta \sigma} c_{\eta k \sigma }^{\dagger } ( 
e^{\dagger} f_{\sigma}^{\pdag}+ \sigma f_{\bar{\sigma}}^{\dagger} d) + {\rm {H.c.}}],
\label{hamiltonian2}
\en
which was proved to be equivalent to the original one (\ref{hamiltonian1}) by Zou and 
Anderson in the case of no spin-flip term.\cite{Zou}

Furthermore, as far as the four possible single electric states are considered as the 
basis, the statistical expectations of the diagonal elements of the density matrix, 
$\rho_{ii}$ ($i=\{0, \sigma, d\}$), give the occupation probabilities of the resonant 
level in the QD being empty, or singly occupied by spin-$\sigma$ electron, or doubly 
occupied, respectively. The nondiagonal term $\rho_{\uparrow \downarrow}$ describes the 
coherent superposition state between the spin-up and -down states in the QD. In the 
slave particle notation, the corresponding relations between the density matrix elements 
and these auxiliary operators are obvious as $\hat \rho_{00}=|0\rangle \langle 0 
|=e^{\dagger} e, \, \hat \rho_{\sigma \sigma}=|\sigma \rangle \langle \sigma 
|=f_{\sigma}^{\dagger} f_{\sigma}^{\pdag}$, $\hat \rho_{dd}=|\uparrow \downarrow \rangle 
\langle \uparrow \downarrow |=d^{\dagger} d^{\pdag}$ and the nondiagonal term $\hat 
\rho_{\sigma \bar{\sigma}}=|\bar{\sigma} \rangle \langle \sigma 
|=f_{\bar{\sigma}}^{\dagger} f_{\sigma}$. According to Eq.~(\ref{comp2}), the constraint 
is subject to the diagonal elements of the density matrix $\hat \rho_{00}+\sum_{\sigma} 
\hat \rho_{\sigma \sigma} + \hat \rho_{dd}=1$.   

\subsection{Derivation of the quantum rate equations}

In this subsection, we derive the rate equations for sequential tunneling starting from 
the combined fermion-boson Hamiltonian (\ref{hamiltonian2}) by using the Keldysh's NGF.

In order to describe the nonequilibrium state of electrons, we define the retarded 
(advanced) and lesser (greater) Green's functions (GFs) for the QD $G_{\sigma 
\sigma'}^{r(a),<(>)}(t,t')\equiv \langle \langle c_{d\sigma}^{\pdag} (t) | 
c_{d\sigma'}^{\dagger}(t')\rangle \rangle^{r(a),<(>)}$ as follows: $G_{\sigma 
\sigma'}^{r(a)}(t,t')\equiv \pm i\theta (\pm t \mp t')\langle \{ c_{d\sigma}^{\pdag} (t) 
, c_{d\sigma'}^{\dagger}(t') \}\rangle$, 
$G_{\sigma \sigma'}^{<}(t,t')\equiv i\langle 
c_{d\sigma'}^{\dagger}(t')c_{d\sigma}^{\pdag} (t) 
\rangle$ and $G_{\sigma \sigma'}^{>}(t,t')\equiv-i\langle c_{d\sigma}^{\pdag} 
(t)c_{d\sigma'}^{\dagger}(t') \rangle$. Considering Eq.~(\ref{sf2}), these GFs in the QD 
can be divided into two parts $G_{\sigma \sigma'}=G_{e \sigma \sigma'}+G_{d \bar{\sigma} 
\bar{\sigma}'}$ with $G_{e \sigma \sigma'}\equiv \langle \langle e^{\dagger}(t) 
f_{\sigma}^{\pdag} (t)| f_{\sigma'}^{\dagger}(t') e(t')\rangle \rangle$ and $G_{d 
\bar{\sigma} \bar{\sigma}'}= \sigma \sigma' \langle \langle f_{\bar 
{\sigma}}^{\dagger}(t) d (t) | d^{\dagger}(t') f_{\bar {\sigma}'}^{\pdag} (t')\rangle 
\rangle$. Under the weak coupling assumption, the central region can be regarded as an 
considerably isolated system and its density matrix operator is supposed to be $\hat 
\rho= \sum_{ij} \rho_{ij} \hat \rho_{ij}$ ($i,j=\{ 0, \sigma, d\}$), meanwhile the 
reservoirs are taken as ``environment" located in local thermal equilibrium. Therefore, 
we can readily define the decoupled diagonal GFs of the QD for weak spin-flip 
transitions in terms of spectrum expression, in the Fourier space, as
\bn
G_{e \sigma \sigma}^{r0}(\omega)&=& \frac{\rho_{00}+\rho_{\sigma \sigma}} {\omega- 
\epsilon_{d}+i0^{+}}, \cr
G_{d \bar{\sigma} \bar{\sigma}}^{r0}(\omega)&=& \frac{\rho_{dd}+\rho_{\bar {\sigma} \bar 
{\sigma}}} {\omega-(\epsilon_{d}+U)+i0^{+}},\cr
G_{e \sigma \sigma}^{<0}(\omega)&=& 2\pi i \rho_{\sigma \sigma} \delta (\omega- 
\epsilon_{d}), \cr
G_{d \bar{\sigma} \bar{\sigma}}^{<0}(\omega) &=& 2\pi i \rho_{dd} \delta [\omega - 
(\epsilon_d +U)], \cr
G_{e \sigma \sigma}^{>0}(\omega) &=& -2\pi i \rho_{00} \delta (\omega - \epsilon_{d}), 
\cr
G_{d \bar{\sigma} \bar{\sigma}}^{>0}(\omega) &=& -2\pi i \rho_{\bar{\sigma} 
\bar{\sigma}} \delta [\omega-(\epsilon_d+U)]. \label{g1}
\en
If no bias voltage is added between the two leads, the central region is in a thermal 
equilibrium state, and the distribution probabilities are well-known as $\rho_{00}=1/Z$, 
$\rho_{\sigma \sigma}=e^{-\epsilon_{d}/T}/Z$, and $\rho_{d d}=e^{-(2\epsilon_d+U)/T}/Z$ 
with $Z=1+2 e^{-\epsilon_{d}/T} + e^{-(2\epsilon_d + U)/T}$. As far as the spin-flip 
transition is considered, the decoupled nondiagonal correlation GFs are crucial in the 
following derivation. Their Fourier expressions are easily obtained from the definitions 
as:
\bn
G_{e \sigma \bar{\sigma}}^{<0}(\omega)&=& 2\pi i \rho_{\sigma \bar{\sigma}} 
\delta(\omega - \epsilon_{d}), \cr
G_{d \bar{\sigma} \sigma}^{>0}(\omega)&=& -2\pi i \rho_{\sigma \bar{\sigma}} 
\delta[\omega - (\epsilon_{d}+U)], \cr
G_{d \bar{\sigma} \sigma}^{<0}(\omega)&=& 0, \quad G_{e \sigma 
\bar{\sigma}}^{>0}(\omega) = 0. \label{g2}
\en
For the case of nonequilibrium interested here, these out-of-equilibrium probabilities 
are determined by the coupling to environments with different chemical potentials, and 
usually they obey a set of equations of time evolution, the rate equations.

Here, we start from the equations of motion of the operators $\hat \rho_{ij}$ with the 
Hamiltonian (\ref{hamiltonian2}) and modified quantization Eq.~(\ref{quan}):
\begin{subequations}
\label{rateq1}
\bq
\dot{\hat \rho}_{00}= i[H, e^{\dagger} e]= -i  \sum_{\eta, k, \sigma} (V_{\eta \sigma} 
c_{\eta k\sigma}^{\dagger} e^{\dagger} f_{\sigma}^{\pdag} - V_{\eta \sigma}^{*} 
f_{\sigma}^{\dagger} e c_{\eta k \sigma}^{\pdag}), \\
\eq
\bn 
\dot{\hat \rho}_{\sigma \sigma}&=& i[H, f_{\sigma}^{\dagger} f_{\sigma}^{\pdag}]
= i \sum_{\eta, k} (V_{\eta \sigma} c_{\eta k\sigma}^{\dagger} e^{\dagger} 
f_{\sigma}^{\pdag} - \bar{\sigma} V_{\eta \bar{\sigma}} c_{\eta k \bar 
{\sigma}}^{\dagger} f_{\sigma}^{\dagger} d \cr 
&& - V_{\eta \sigma}^{*} f_{\sigma}^{\dagger} e c_{\eta k \sigma}^{\pdag} + \bar{\sigma} 
V_{\eta \bar{\sigma}}^{*} d^{\dagger} f_{\sigma}^{\pdag} c_{\eta k \bar{\sigma}}) \cr 
&& + iR_{\bar{\sigma}} f_{\bar{\sigma}}^{\dagger} f_{\sigma}^{\pdag} - iR_{\sigma} 
f_{\sigma}^{\dagger} f_{\bar{\sigma}}^{\pdag} , \label{rateq1-2} \\
\dot{\hat \rho}_{\sigma \bar{\sigma}}&=& i[H, f_{\bar{\sigma}}^{\dagger} 
f_{\sigma}^{\pdag}]
= i \sum_{\eta, k} (V_{\eta \bar{\sigma}} c_{\eta k \bar{\sigma}}^{\dagger} e^{\dagger} 
f_{\sigma}^{\pdag} - \bar{\sigma} V_{\eta \bar{\sigma}} c_{\eta k \bar 
{\sigma}}^{\dagger} f_{\bar{\sigma}}^{\dagger} d \cr 
&& - V_{\eta \sigma}^{*} f_{\bar{\sigma}}^{\dagger} e c_{\eta k \sigma}^{\pdag} + \sigma 
V_{\eta \sigma}^{*} d^{\dagger} f_{\sigma}^{\pdag} c_{\eta k \sigma}) \cr 
&& + iR_{\sigma} (f_{\sigma}^{\dagger}f_{\sigma}^{\pdag} - f_{\bar{\sigma}}^{\dagger} 
f_{\bar{\sigma}}^{\pdag} ),
\en
\bq
\dot {\hat \rho}_{dd} = i[H, d^{\dagger}d]=i \sum_{\eta,k,\sigma} \sigma ( V_{\eta 
\sigma} c_{\eta k\sigma}^{\dagger} f_{\bar{\sigma}}^{\dagger} d- V_{\eta \sigma}^{*} 
d^{\dagger} f_{\bar{\sigma}}^{\pdag} c_{\eta k\sigma}). 
\eq
\end{subequations}
Their statistical expectations involve the time-diagonal parts of the correlation 
functions: $G_{e\sigma,\eta k \sigma'}^{<}(t,t')\equiv i\langle c_{\eta 
k\sigma'}^{\dagger}(t') e^{\dagger}(t) f_{\sigma}^{\pdag}(t)\rangle$, $G_{d \sigma',\eta 
k \sigma}^{<}(t,t')\equiv i\langle c_{\eta k\sigma}^{\dagger}(t') 
f_{\sigma'}^{\dagger}(t) d(t)\rangle$, $G_{\eta k\sigma',e \sigma}^{<}(t,t')\equiv 
i\langle 
f_{\sigma}^{\dagger}(t')e(t') c_{\eta k\sigma'}^{\pdag}(t)\rangle$, and $G_{\eta 
k\sigma,d \sigma'}^{<}(t,t')\equiv i\langle 
d^{\dagger}(t') f_{\sigma'}^{\pdag}(t') c_{\eta k\sigma}^{\pdag}(t)\rangle$. With the 
help of the Langreth analytic continuation rules, \cite{Langreth} we obtain the 
following expressions in the wide band limit (The detail derivation will be given in the 
Appendix):
\begin{widetext} 
\begin{subequations}
\label{rateq2}
\bq
\dot{\rho}_{00}= -\frac{i}{2\pi} \int d\omega \sum_{\eta, \sigma} \{ \Gamma_{\eta 
\sigma}f_{\eta}(\omega)G_{e \sigma\sigma}^{>}(\omega)
+ \Gamma_{\eta \sigma} [1-f_{\eta}(\omega)] G_{e \sigma\sigma}^{<}(\omega) \}, \eq
\bn
\dot{\rho}_{\sigma \sigma}&=& \frac{i}{2\pi} \int d\omega \sum_{\eta} \{ \Gamma_{\eta 
\sigma}f_{\eta}(\omega)G_{e \sigma\sigma}^{>}(\omega)
+ \Gamma_{\eta \sigma} [1-f_{\eta}(\omega)] G_{e \sigma\sigma}^{<}(\omega) 
- \Gamma_{\eta \bar{\sigma}} f_{\eta}(\omega) G_{d \sigma\sigma}^{>}(\omega) - 
\Gamma_{\eta \bar{\sigma}} [1-f_{\eta}(\omega)] G_{d \sigma \sigma}^{<}(\omega) \} \cr 
&& + iR_{\bar{\sigma}} \rho_{\sigma \bar{\sigma}} - iR_{\sigma} \rho_{\bar{\sigma} 
\sigma}, \label{rateq2-2}
\en
\bn
\dot{\rho}_{\sigma \bar{\sigma}}&=& \frac{i}{4\pi} \int d\omega \sum_{\eta} 
(\Gamma_{\eta \sigma}+ \Gamma_{\eta \bar{\sigma}}) \{ f_{\eta}(\omega) G_{e \sigma 
\bar{\sigma}}^{>}(\omega)
+ [1-f_{\eta}(\omega)] G_{e \sigma \bar{\sigma}}^{<}(\omega) - f_{\eta}(\omega) G_{d 
\bar{\sigma} \sigma}^{>}(\omega) - [1-f_{\eta}(\omega)] G_{d \bar{\sigma} 
\sigma}^{<}(\omega) \} \cr
&& + iR_{\sigma} (\rho_{\sigma \sigma}-\rho_{\bar{\sigma}\bar{\sigma}}),
\en
\bq
\dot{\rho}_{dd}= \frac{i}{2\pi} \int d\omega \sum_{\eta, \sigma} \{ \Gamma_{\eta \sigma} 
f_{\eta}(\omega) G_{d \bar{\sigma} \bar{\sigma}}^{>}(\omega) + \Gamma_{\eta \sigma} 
[1-f_{\eta}(\omega)] G_{d \bar{\sigma} \bar{\sigma}}^{<} \}, 
\eq
\end{subequations}
\end{widetext}
in terms of the QD's GFs in the Fourier space.\cite{Meir} Here $\Gamma_{\eta 
\sigma}=2\pi \sum_{k}|V_{\eta \sigma}|^2 \delta (\omega-\epsilon_{\eta k\sigma})$ 
denotes the strength of coupling between the QD level and the lead $\eta$. In wide band 
limit, it is independent of energy and is supposed to be constant. Under the weak 
coupling assumption, it is adequate to keep only the leading order of $|V|^2$ in 
evaluation of these occupation densities. So we can replace these QD's GFs with their 
decoupled formulas Eqs.~(\ref{g1}) and (\ref{g2}). Finally, the resulting quantum rate 
equations become:
\begin{subequations}
\label{rateqSQD}
\bq
\dot{\rho}_{00}= \sum_{\sigma} ( \Gamma_{\sigma}^{-} \rho_{\sigma \sigma} - 
\Gamma_{\sigma}^{+} \rho_{00}), \label{r0}
\eq
\bq
\dot{\rho}_{\sigma \sigma}= \Gamma_{\sigma}^{+} \rho_{00}+\widetilde 
{\Gamma}_{\bar{\sigma}}^{-} \rho_{dd}- ( \Gamma_{\sigma}^{-} + \widetilde 
{\Gamma}_{\bar{\sigma}}^{+}) \rho_{\sigma \sigma} - 2\Im (R_{\bar{\sigma}}\rho_{\sigma 
\bar{\sigma}}), \label{r1}
\eq
\bq
\dot{\rho}_{\sigma \bar{\sigma}} = iR_{\sigma}(\rho_{\sigma \sigma} - \rho_{\bar{\sigma} 
\bar{\sigma}})  - \frac{1}{2} ( \widetilde\Gamma_{\sigma}^{+} + 
\widetilde\Gamma_{\bar{\sigma}}^{+} + \Gamma_{\sigma}^{-} + \Gamma_{\bar{\sigma}}^{-}) 
\rho_{\sigma \bar{\sigma}}, \label{r2}
\eq
\bq
\dot{\rho}_{dd}= \widetilde{\Gamma}_{\downarrow}^{+} \rho_{\uparrow \uparrow} + 
\widetilde {\Gamma}_{\uparrow}^{+} \rho_{\downarrow \downarrow} - ( 
\widetilde{\Gamma}_{\uparrow}^{-} + \widetilde{\Gamma}_{\downarrow}^{-}) \rho_{dd}, 
\label{r3}       
\eq
\end{subequations}
together with the normalization relation $\rho_{00}+ \rho_{dd}+ \sum_{\sigma} 
\rho_{\sigma \sigma}=1$ from Eq.~(\ref{comp}), with the definitions 
$\Gamma_{\sigma}^{\pm}= \sum_{\eta} \Gamma_{\eta \sigma} f_{\eta}^{\pm}(\epsilon_{d})$ 
and $\widetilde{\Gamma}_{\sigma}^{\pm}= \sum_{\eta} \Gamma_{\eta \sigma} 
f_{\eta}^{\pm}(\epsilon_{d}+U)$, where $f_{\eta}^{+}(\omega)=\{1+ 
e^{(\omega-\mu_{\eta})/T} \}^{-1}$ is the Fermi distribution function of the $\eta$ lead 
and $f^{-}(\omega)=1-f^{+}(\omega)$. Here, $\Gamma_{\sigma}^{+}$ ($\Gamma_{\sigma}^{-}$) 
describes the tunneling rate of electrons with spin-$\sigma$ into (out from) the QD 
without the occupation of the $\bar{\sigma}$ state. Similarly, $\widetilde 
{\Gamma}_{\sigma}^{+}$ ($\widetilde {\Gamma}_{\sigma}^{-}$) describes the tunneling rate 
of electrons with spin-$\sigma$ in to (out from) the QD, when the QD is already occupied 
by an electron with spin-$\bar{\sigma}$, revealing the modification of the corresponding 
rates due to the Coulomb repulsion.   

These rate equations Eqs.~(\ref{r0}), (\ref{r1}) and (\ref{r3}) coincide with the 
previous classical rate equations in the sequential picture for the resonant tunneling 
if the intradot spin-flip transition is quenched.\cite{Glazman,Beenakker} Obviously, if 
the left lead has the same chemical potential as the right lead, the stationary 
solutions of Eqs.~(\ref{r0}), (\ref{r1}) and (\ref{r3}) reduce exactly to the 
above-mentioned thermal equilibrium results in the case of $R=0$. In this situation, 
they have clear classical interpretations. For example, the rate of change of the number 
of the spin-$\sigma$ electrons $\rho_{\sigma \sigma}$ in the SQD, described by 
Eq.~(\ref{r1}), is contributed from the following four single-particle tunneling 
processes: 1) tunneling into the QD with spin-$\sigma$ electrons $\Gamma_{\sigma}^{+}$ 
from both left and right leads if the QD is initially in the empty state $\rho_{00}$; 2) 
tunneling out from the QD with spin-$\bar{\sigma}$ electrons $\widetilde 
{\Gamma}_{\bar{\sigma}}^{-}$ into both two leads if the QD is initially in the doubly 
occupied state $\rho_{dd}$; 3) tunneling into the QD with spin-$\bar{\sigma}$ electrons 
$\widetilde {\Gamma}_{\bar{\sigma}}^{+}$ from both two leads; and 4) tunneling out from 
the QD with spin-$\sigma$ electrons $\Gamma_{\sigma}^{-}$ into both two leads, when the 
QD is initially just in this state $\rho_{\sigma \sigma}$. Tunneling events 1) and 2) 
increase the probability of the spin-$\sigma$ state, but events 3) and 4) decrease this 
probability. These contributions constitute the classical rate equation form. Other 
diagonal equations have similar interpretations. Notice that the final term in 
Eq.~(\ref{r1}) describes transitions between isolated states through the coupling with 
nondiagonal terms, which has no classical counterpart. Therefore, it is responsible for 
coherent effects in the transport.              

The nondiagonal matrix element $\rho_{\sigma \bar{\sigma}}$ is ruled by Eq.~(\ref{r2}), 
which resembles the optical Bloch equation and describes the dynamics of quantum 
superposition. This is a pure quantum effect. As mentioned by Gurvitz and 
Prager,\cite{Gurvitz} the couplings with the leads (all possible tunneling processes 
involved) always provide negative contribution and cause damping of the quantum 
superposition.

The particle current $I_{\eta}$ flowing from the lead $\eta$ to the QD can be evaluated 
from the rate of time change of the electron number operator $N_{\eta}(t)=\sum_{k, 
\sigma} c_{\eta k\sigma}^{\dagger}(t) c_{\eta
k\sigma}^{\pdag}(t)$ of the lead $\eta$:\cite{Meir}
\bn
{I}_{\eta}(t)&=& -\frac{e}{\hbar}\Big\langle \frac{d{N}_{\eta}}{dt}\Big\rangle=
-i\frac{e}{\hbar}\Big\langle \Big [H, \sum_{k,\sigma}c_{\eta k \sigma}^{\dagger}(t)
c_{\eta k \sigma}(t)\Big ] \Big\rangle \cr 
&=& i\frac{e}{\hbar} \Big\langle \sum_{k,\sigma}
\{ V_{\eta \sigma}^{\pdag} c_{\eta k \sigma}^\dagger(t) [ e^{\dagger}(t) 
f_{\sigma}^{\pdag}(t) +   \sigma f_{\bar{\sigma}}^{\dagger}(t) d(t) ] \cr
&& - V_{\eta \sigma}^{*} [ f_{\sigma}^\dagger(t)e(t) + \sigma d^{\dagger} 
f_{\bar{\sigma}}(t) ] c_{\eta k\sigma}^{\pdag}(t) \} \Big\rangle. \label{i}
\en
Ultimately, the current can be expressed in terms of the GFs in the QD:
\bn
I_{\eta}&=& ie\int \frac{d\omega}{2\pi} \sum_{\sigma} \{ \Gamma_{\eta \sigma} 
f_{\eta}(\omega) [G_{e\sigma\sigma}^{>}(\omega) + G_{d \bar{\sigma} 
\bar{\sigma}}^{>}(\omega)] + \cr 
&& \Gamma_{\eta \sigma} [1-f_{\eta}(\omega)] [G_{e\sigma\sigma}^{<}(\omega) + 
G_{d\bar{\sigma} \bar{\sigma}}^{<}] \}. \label{ii}  
\en
Under the weak coupling approximation, it becomes
\bq
I_{\eta}/e=\sum_{\sigma} ( \widetilde{\Gamma}_{\eta \sigma}^{-} \rho_{dd} + \Gamma_{\eta 
\sigma}^{-} \rho_{\sigma \sigma} - \widetilde{\Gamma}_{\eta \bar{\sigma}}^{+} 
\rho_{\sigma \sigma} - \Gamma_{\eta \sigma}^{+} \rho_{00}). \label{iii}
\eq 
This formulae demonstrates that all possible tunneling processes relevant to the lead 
$\eta$ can provide corresponding contributions to the current of the lead $\eta$ and the 
current is totally determined by the diagonal elements of the density matrix of the 
central region. However, the nondiagonal element of the density matrix is coupled with 
diagonal elements in the rate equation Eq.~(\ref{r1}), and therefore influence the 
tunneling current indirectly. 

\subsection{Discussions}

The rate equations (\ref{rateqSQD}) may be readily solved under stationary condition for 
arbitrary bias voltages $V$ and temperatures $T$, and consequently the dc current may be 
obtained via Eq.~(\ref{iii}). More interestingly, it is useful to review the following 
two special cases in the case of large Coulomb repulsion. First, we consider that no 
doubly occupied state is available in the QD, i.e., $\rho_{dd}=0$. In this case, we 
assume that the bare-level $\epsilon_{d}$ of the QD is just above the Fermi levels $\mu$ 
of the two leads under equilibrium condition, meaning 
$\widetilde{\Gamma}_{\sigma}^{+}\simeq 0$, $\widetilde{\Gamma}_{\sigma}^{-} \simeq 
\sum_{\eta} \Gamma_{\eta \sigma}$. Then, in steady state, the quantum rate equations 
(\ref{r1}) and (\ref{r2}) become
\begin{subequations}
\bn
&&\Gamma_{\sigma}^{+} \rho_{00}- \Gamma_{\sigma}^{-} \rho_{\sigma \sigma} - 2\Im 
(R_{\bar{\sigma}} \rho_{\sigma \bar{\sigma}})=0, \\
&&iR_{\sigma}(\rho_{\sigma \sigma} - \rho_{\bar{\sigma} \bar{\sigma}})- \frac{1} {2} 
(\Gamma_{\sigma}^{-} + \Gamma_{\bar{\sigma}}^{-}) \rho_{\sigma \bar{\sigma}}=0,
\en
\end{subequations}
with $\rho_{00}+\sum_{\sigma} \rho_{\sigma \sigma}=1$. They can be readily solved
\begin{subequations}
\label{OO1}
\bn
\rho_{\uparrow \uparrow}&=& \frac{\Gamma_{\uparrow}^{+} \Gamma_{\downarrow}^{-} + \chi 
(\Gamma_{\uparrow}^{+}+ \Gamma_{\downarrow}^{+})} {\Delta}, \label{o1} \\ 
\rho_{\downarrow \downarrow}&=& \frac{\Gamma_{\downarrow}^{+} \Gamma_{\uparrow}^{-} + 
\chi (\Gamma_{\uparrow}^{+} + \Gamma_{\downarrow}^{+})} {\Delta}, \label{o2}
\en
\end{subequations}
in which $\Delta=[(\Gamma_{\uparrow}^{+} + \Gamma_{\uparrow}^{-}) 
(\Gamma_{\downarrow}^{+} + \Gamma_{\downarrow}^{-})- \Gamma_{\uparrow}^{+} 
\Gamma_{\downarrow}^{+}] + \chi (2\Gamma_{\uparrow}^{+}+2\Gamma_{\downarrow}^{+} + 
\Gamma_{\uparrow}^{-} + \Gamma_{\downarrow}^{-})$ and $\chi=4 
|R|^2/(\Gamma_{\uparrow}^{-} + \Gamma_{\downarrow}^{-})$. The steady tunneling current 
is $I_{R}=e\sum_{\sigma} [(\Gamma_{R \sigma}^{+} + \Gamma_{R \bar{\sigma}}^{+} + 
\Gamma_{R \sigma}^{-}) \rho_{\sigma \sigma} 
- \Gamma_{R \sigma}^{+}]$. For large bias voltage, i.e., $\Gamma_{L\sigma}^{-}=0$ and 
$\Gamma_{R\sigma}^{+}=0$, and spin-independent tunneling, the dc current becomes
\bq
I_{R}/e=\frac{2\Gamma_{L} (\Gamma_{R} + 2\chi')}{2\Gamma_{L} + \Gamma_{R} + \chi' (2+ 4 
\Gamma_{L}/\Gamma_{R})},
\eq
and $\chi'=2|R|^2/\Gamma_{R}$, which depicts the spin-flip transition induced 
modification for the corresponding formula Eq.~(3.10) in Ref.~\onlinecite{Gurvitz}. 

The second case we considered here is the deep level in the large-$U$ limit: the 
bare-level $\epsilon_d$ is far below the Fermi level $\mu$ but $\epsilon_{d}+U$ is 
slightly above the Fermi level in equilibrium condition, implicating that the QD is 
always occupied by electrons. In this situation, we have $\rho_{00}=0$, 
$\Gamma_{\sigma}^{-}\simeq 0$, and $\Gamma_{\sigma}^{+} \simeq \sum_{\eta} \Gamma_{\eta 
\sigma}$. Different from the above case, only singly and doubly occupied states are 
permitted in tunneling processes. Under the stationary condition, the quantum rate 
equations (\ref{r1}) and (\ref{r2}) reduce to
\begin{subequations}
\bn
&& \widetilde{\Gamma}_{\bar{\sigma}}^{-} \rho_{dd} - 
\widetilde{\Gamma}_{\bar{\sigma}}^{+} \rho_{\sigma \sigma} - 2\Im (R_{\bar{\sigma}} 
\rho_{\sigma \bar{\sigma}})=0, \\
&& iR_{\sigma}(\rho_{\sigma \sigma} - \rho_{\bar{\sigma} \bar{\sigma}}) - \frac{1}{2} 
(\widetilde{\Gamma}_{\sigma}^{+} + \widetilde{\Gamma}_{\bar{\sigma}}^{+}) \rho_{\sigma 
\bar{\sigma}}=0,
\en
\end{subequations}
with $\rho_{dd}+\sum_{\sigma} \rho_{\sigma \sigma}=1$. The solutions are
\begin{subequations}
\label{OO2}
\bq
\rho_{\uparrow \uparrow}= \left [ \frac{(\widetilde{\Gamma}_{\uparrow}^{+} + \chi) 
(\widetilde{\Gamma}_{\downarrow}^{+} + \widetilde{\Gamma}_{\downarrow}^{-}) + \chi 
(\widetilde{\Gamma}_{\uparrow}^{+} + \widetilde{\Gamma}_{\uparrow}^{-})} 
{(\widetilde{\Gamma}_{\uparrow}^{+} + \chi) (\widetilde{\Gamma}_{\downarrow}^{+} + \chi) 
- \chi^2} - 1 \right ] \rho_{dd}, \label{o3}
\eq
\bq
\rho_{\downarrow \downarrow}= \left [ \frac{(\widetilde{\Gamma}_{\downarrow}^{+} + \chi) 
(\widetilde{\Gamma}_{\uparrow}^{+} + \widetilde{\Gamma}_{\uparrow}^{-}) + \chi 
(\widetilde{\Gamma}_{\downarrow}^{+} + \widetilde{\Gamma}_{\downarrow}^{-})} 
{(\widetilde{\Gamma}_{\uparrow}^{+} + \chi) (\widetilde{\Gamma}_{\downarrow}^{+} + \chi) 
- \chi^2} - 1 \right ] \rho_{dd}, \label{o4}
\eq
\bn
\rho_{dd}&=& \left [ \frac{(\widetilde{\Gamma}_{\downarrow}^{+} + \chi) 
(\widetilde{\Gamma}_{\uparrow}^{+} + \widetilde{\Gamma}_{\uparrow}^{-}) + \chi 
(\widetilde{\Gamma}_{\downarrow}^{+} + \widetilde{\Gamma}_{\downarrow}^{-})} 
{(\widetilde{\Gamma}_{\downarrow}^{+} + \chi) \widetilde{\Gamma}_{\uparrow}^{+} + \chi 
\widetilde{\Gamma}_{\downarrow}^{+}} \right. \cr
&& + \left. \frac{(\widetilde{\Gamma}_{\uparrow}^{+} + \chi) 
(\widetilde{\Gamma}_{\downarrow}^{+} + \widetilde{\Gamma}_{\downarrow}^{-}) + \chi 
(\widetilde{\Gamma}_{\uparrow}^{+} + \widetilde{\Gamma}_{\uparrow}^{-})} 
{(\widetilde{\Gamma}_{\uparrow}^{+} + \chi) \widetilde{\Gamma}_{\downarrow}^{+} + \chi 
\widetilde{\Gamma}_{\uparrow}^{+}} -1 \right ]^{-1},\cr
&& \label{o5}
\en
\end{subequations}
with $\chi=4|R|^2/(\widetilde{\Gamma}_{\uparrow}^{+} + 
\widetilde{\Gamma}_{\downarrow}^{+})$. The steady tunneling current is given by 
$I_{R}=e\sum_{\sigma} [\widetilde{\Gamma}_{R\sigma}^{-} \rho_{dd} - 
\widetilde{\Gamma}_{R \bar{\sigma}}^{+} \rho_{\sigma \sigma}]$. Large bias voltage 
further simplifies the spin-independent current as
\bq
I_{R}/e=\frac{2\widetilde{\Gamma}_{L} \widetilde{\Gamma}_{R} + 2\chi^{\prime\prime}} 
{\widetilde{\Gamma}_{L} + 2 \widetilde{\Gamma}_{R} + 2\chi^{\prime\prime} (1+ 
\widetilde{\Gamma}_{R}/\widetilde{\Gamma}_{L})},
\eq 
with $\chi^{\prime\prime}=2|R|^2/\widetilde{\Gamma}_{L}$. This is a modification of 
Eq.~(3.11) given by Gurvitz and Prager due to spin-flip transitions.\cite{Gurvitz}

It should be noted that the same two cases are also analyzed in 
Ref.~\onlinecite{Rudzinski} for an interacting QD with spin-flip transitions included. 
They evaluated the occupation numbers from the classical rate equations and utilized a 
spin relaxation time $\tau_s$ to describe the spin-flip transitions. Therefore, their 
results are slightly different from ours. For both cases, if we redefine the spin 
relaxation time $\tau_s$ as $1/\tau_s=\chi$, which is now of temperature and bias 
voltage dependence, their results\cite{Rudzinski} are the same as ours, Eqs.~(\ref{OO1}) 
and (\ref{OO2}) of the quantum rate equations. This is a clear demonstration of the 
importance of quantum ``coherence".  

It is also worth examining the quantum rate equations (\ref{rateqSQD}) derived here at 
large bias voltage between the left and right leads without the spin-flip transitions. 
In this case, we assume $eV\gg T$ and $eV\gg U$, so $\Gamma_{L \sigma}^{-}=\Gamma_{R 
\sigma}^{+}=0$ and $\widetilde\Gamma_{L \sigma}^{-}=\widetilde\Gamma_{R \sigma}^{+}=0$. 
The quantum rate equations lead:
\begin{subequations}
\bn
\dot{\rho}_{00}&=& \Gamma_{R \uparrow}\rho_{\uparrow \uparrow} + \Gamma_{R \downarrow} 
\rho_{\downarrow \downarrow} - (\Gamma_{L \uparrow} + \Gamma_{L \downarrow}) \rho_{00}, 
\\
\dot{\rho}_{\sigma \sigma}&=& \Gamma_{L \sigma} \rho_{00} + \widetilde{\Gamma}_{R 
\bar{\sigma}} \rho_{dd} - (\Gamma_{R \sigma} + \widetilde{\Gamma}_{L \bar{\sigma}}) 
\rho_{\sigma \sigma}, \\
\dot{\rho}_{dd}&=& \widetilde{\Gamma}_{L \downarrow} \rho_{\uparrow \uparrow} + 
\widetilde{\Gamma}_{L \uparrow} \rho_{\downarrow \downarrow} - (\widetilde{\Gamma}_{R 
\uparrow}+ \widetilde{\Gamma}_{R \downarrow}) \rho_{dd}.             
\en
\end{subequations}
And the current is $I_{R}=e\sum_{\sigma} (\widetilde{\Gamma}_{R \sigma} \rho_{dd} + 
\Gamma_{R \sigma} \rho_{\sigma \sigma})$. At zero temperature and spin-independent 
tunneling, these equations indeed resemble the rate equations derived from the 
Schr\"odinger equation developed by Gurvitz and Prager.\cite{Gurvitz} 

\subsection{Numerical results}

In this subsection, we perform numerical calculations for the spin dependence of the 
tunneling processes, through the SQD connected to two ferromagnetic leads. In the 
following calculations, we consider two magnetic configurations, namely, parallel (P) 
and antiparallel (AP) configurations. When the magnetic electrodes are in P 
configuration, we assume that the majority electrons are spin-up $\sigma=\uparrow$, and 
the minority electrons are spin-down $\sigma=\downarrow$. We also assume that in the AP 
configuration the magnetization of the right electrode is reversed. 

Therefore, for the identical leads and symmetric barriers, of interest in the present 
paper, we further assume that the ferromagnetism of the leads can be accounted for by 
the polarization-dependent couplings 
$\Gamma_{L\uparrow}=\Gamma_{R\uparrow}=(1+p)\Gamma_{0}$, 
$\Gamma_{L\downarrow}=\Gamma_{R\downarrow}=(1-p)\Gamma_{0}$ for the P alignment, while 
$\Gamma_{L\uparrow}=\Gamma_{R\downarrow}=(1+p)\Gamma_{0}$, 
$\Gamma_{L\downarrow}=\Gamma_{R\uparrow}=(1-p)\Gamma_{0}$ for the AP alignment. 
$\Gamma_{0}$ denotes the tunneling coupling between the QD and the leads without 
internal magnetization, and $p$ ($0\leq p< 1$) stands for the polarization strength of 
the leads. In the wide band limit, $\Gamma_0$ is supposed to be a constant and chosen as 
unit of energy in the following paper. Moreover, we measure energy from the Fermi levels 
of the left and right leads in the equilibrium condition ($\mu_{L}=\mu_{R}=0$) 
thereafter. The discrete level $\epsilon_d$ of the QD can cross the Fermi levels by 
tuning the gate voltage in experiments. Without loss of generality, we apply the bias 
voltage $V$ between the source and drain symmetrically $\mu_{L}=-\mu_{R}=eV/2$, and 
neglect the shift of the discrete level caused by this external voltage. Because of the 
symmetry, we will restrict to positive bias only, $V>0$. 

From Eqs.~(\ref{rateqSQD}), one can find all the expectation values of the 
density-matrix elements for a given bias $V$ in the stationary condition, and thus allow 
us to calculate the tunneling current flowing through the system by employing 
Eq.~(\ref{iii}) and the nonequilibrium occupation numbers $n_{\uparrow}$, 
$n_{\downarrow}$, defined by
$n_{\sigma}=\rho_{\sigma \sigma}+\rho_{dd}$.

First we consider no spin-flip scattering processes on the QD. Figs.~2(a) and (b) plot 
the nonequilibrium occupation numbers as a function of the bare level calculated for a 
small bias $V=1.0$ and a large bias $V=10.0$, respectively, in both P (thin lines) and 
AP (thick lines) configurations. The two spacial bias voltages are chosen in order here 
to demonstrate the linear response regime ($V=1.0$) and the strong nonlinear case 
($V=10.0$), respectively. For comparison, we also plot the equilibrium occupation 
numbers in Fig.~2(b). From these figure, we can observe that: 1) The complete Coulomb 
blockade (charging) effect in equilibrium (the single step in $\rho_{dd}$) is partially 
removed in nonequlibrium, i.e., $\rho_{dd}$ becomes a multi-step function of the gate 
voltage; 2) $n_{\sigma}$ has fractional steps in nonequilibrium in contrast to just 
half-interger steps in equilibrium; 3) $n_{\uparrow}=n_{\downarrow}$ in the P 
configuration, whereas $n_{\uparrow}\neq n_{\downarrow}$ in the AP configuration. 
Fig.~2(c) shows the tunneling current calculated for both configurations. The current in 
the P alignment is always larger than that in the AP alignment in the whole range of the 
gate voltage. In the linear response regime, the current provides the information of the 
conductance of the device: there appear two resonant peaks with equal heights when the 
gate voltage controlled levels $\epsilon_d$ and $\epsilon_d+U$ respectively cross the 
Fermi levels of two leads. While in the strong nonequilibrium case, there are three 
steps in the current, which correspond to the steps in the occupation numbers, whereas 
between the steps the current is constant. 

Fig.~3 illustrates typical variations of the occupation numbers and the current with the 
bias voltage $V$ for $\epsilon_{d}=1$ (the no doubly occupied level) and 
$\epsilon_{d}=-5$ (the deep level). In the first (second) case, the first step in 
$n_{\sigma}$ occurs at the bias, when the Fermi level of the source or drain crosses the 
discrete level $\epsilon_{d}$ ($\epsilon_d+U$). This means a new channel open for 
tunneling. Consequently, we find a step in the current appears at this position. As the 
bias further increases, they all keep constant until the second step at a higher voltage 
corresponding to the case when the Fermi level crosses $\epsilon_{d}+U$ ($\epsilon_d$), 
which also induces a step in $\rho_{dd}$. The insets in Fig.~3(c) and (f) depict the 
corresponding tunnel magnetoresistance (TMR), defined as 
\bq
\text{TMR}\equiv \frac{I_{\text{P}}-I_{\text{AP}}}{I_{\text{AP}}}. 
\eq
The TMR is enhanced by the Coulomb interaction in the range between the two biases 
corresponding to the two steps in the current. In these figures, we also display the 
temperature effect in tunneling characteristics. It is easily observed that increasing 
temperature gradually smoothes the steep step structure in the occupation numbers and 
the current, and decreases the TMR.

We now consider the effect of spin-flip scatterings on the tunneling. Because the 
spin-flip processes have no influence on the occupation numbers and the current in the P 
configuration, we plot the calculated results for the AP configuration with $R=1$ in 
Fig.~4. It is obvious, in comparison with the case of no spin-flip scattering $R=0$ 
(thin lines), that the spin-flip transition decreases the difference between 
$n_{\uparrow}$ and $n_{\downarrow}$, increases $\rho_{dd}$ and the current. Moreover, 
their temperature behaviors are similar with the case of no spin-flip transition. It is 
worth noting that when the bias voltage is lower than $10.0$, i.e., the value 
corresponding to the second step in $n_{\sigma}$ and the only step in $\rho_{dd}$, we 
have approximately $\rho_{dd}\simeq 0$ (no doubly occupied level) for $\epsilon_{d}=1$ 
and $\rho_{00}\simeq 0$ (deep level) for $\epsilon_{d}=-5$, indicating that 
Eqs.~(\ref{OO1}) and (\ref{OO2}) are valid in this bias range. Therefore, we can utilize 
the definition of the spin relaxation rate in these equations to account for the 
importance of temperature and bias on the spin-flip scattering, as depicted in the 
insets of Figs.~4(b) and (e).  

\section{Coupled quantum dots}

Now we turn to resonant tunneling through a CQD with weak coupling between the QDs and 
the leads, as shown in Fig.~1(b). The presumption that the interdot hopping is also weak 
keeps each level of the dots isolated. Then the superposition of the two levels in 
different QDs plays a crucial role in tunneling. In order to simplify our derivation, we 
consider here the infinite intradot Coulomb repulsion $U'$ and a finite interdot Coulomb 
interaction $U$, which excludes the state of two electrons in the same QD but two 
electrons can occupy different QDs.        

\subsection{Model Hamiltonian and slave-particle representation}

The tunneling Hamiltonian for the CQD is
\bn
H &=& \sum_{\eta, k, \sigma}\epsilon _{\eta k\sigma} 
c_{\eta k\sigma }^{\dagger }c_{\eta k\sigma }^{\pdag}+ \epsilon_{1} \sum_{\sigma} c_{1 
\sigma }^{\dagger }c_{1 \sigma }^{\pdag}+ \epsilon_{2} \sum_{\sigma} c_{2 \sigma 
}^{\dagger }c_{2 \sigma }^{\pdag} \cr
&& + t\sum_{\sigma} (c_{1 \sigma}^{\dagger} c_{2\sigma}^{\pdag} + c_{2 \sigma}^{\dagger} 
c_{1\sigma}^{\pdag})
+ U'n_{1 \uparrow }n_{1 \downarrow }+ U'n_{2 \uparrow }n_{2 \downarrow } \cr
&& + U \sum_{\sigma, \sigma'} n_{1 \sigma}n_{2 \sigma'}  +\sum_{k, \sigma} (V_{L \sigma} 
c_{L k\sigma }^{\dagger }c_{1 \sigma}^{\pdag} +{\rm {H.c.}}) \cr && + \sum_{k, \sigma} 
(V_{R \sigma} c_{R k\sigma }^{\dagger }c_{2 \sigma}^{\pdag} +{\rm {H.c.}}),
\label{hamiltonian3}
\en
where $c_{1(2)\sigma}^{\dagger}$, $c_{1(2)\sigma}$ are creation and annihilation 
operators for a spin-$\sigma$ electron in the first (second) QD, respectively. 
$\epsilon_{j}$ ($j=1,2$) is the bare-level energy of electrons in the $j$th QD, 
$\epsilon_{1(2)}=\epsilon_{d} \pm \delta$, in which $\delta$ is the bare mismatch 
between the two bare levels. The first term in the second line denotes the hopping $t$ 
between the two QDs. The other notations are the same with those in Sec. II.

In the situation interested here, the bare mismatch $\delta$ should be very small. 
Otherwise, the quantum coherence (the superposition of the two states) has quite tiny 
effect on the tunneling processes. In experiments, this small mismatch could be 
controlled by external time-dependent voltages. The available states and the 
corresponding energies for the isolated CQD are: (1) the whole system is empty, 
$|0\rangle_{1} |0\rangle_{2} $, and the energy is zero; (2) the first QD is singly 
occupied, $|\sigma\rangle_{1} |0\rangle_{2}$, and the energy is $\epsilon_{1}$; (3) the 
second QD is singly occupied, $|0\rangle_{1} |\sigma\rangle_{2}$, and the energy is 
$\epsilon_2$; (4) both of the QDs are singly occupied, $|\sigma\rangle_{1} 
|\sigma'\rangle_{2}$, and the energy is $2\epsilon_{d} + U$. With the same theoretical 
point of view as in the single QD mentioned in the above section, we can decompose the 
real electron operator $c_{j\sigma}$ in these Fock states as
\bn
c_{1\sigma} &=& |0\rangle_{1} |0\rangle_{2} {}_{2}\langle 0| {}_{1}\langle \sigma| + 
\sum_{\sigma'} |0\rangle_{1} |\sigma'\rangle_{2} {}_{2}\langle\sigma'| {}_{1}\langle 
\sigma|, \label{sf3} \\
c_{2\sigma} &=& |0\rangle_{1} |0\rangle_{2} {}_{2}\langle 0| {}_{1}\langle \sigma| + 
\sum_{\sigma'} |\sigma'\rangle_{1} |0\rangle_{2} {}_{2}\langle\sigma| {}_{1}\langle 
\sigma'|, \label{sf4}
\en
in association with the completeness relation
\bn
&&|0\rangle_{1} |0\rangle_{2} {}_{2}\langle 0| {}_{1}\langle 0| + \sum_{\sigma, \sigma'} 
|\sigma\rangle_{1} |\sigma'\rangle_{2} {}_{2}\langle\sigma'| {}_{1}\langle\sigma| \cr
&& + \sum_{\sigma} (|\sigma\rangle_{1} |0\rangle_{2} {}_{2}\langle 0| \langle \sigma| + 
|0\rangle_{1} |\sigma\rangle_{2} {}_{2}\langle \sigma| {}_{1}\langle 0|)
=1. \label{comp3}
\en
Again, we assign these Dirac brackets as operators: the slave-boson operators 
$e^{\dagger}=|0\rangle_{1} |0\rangle_{2}$, $d_{\sigma 
\sigma'}^{\dagger}=|\sigma\rangle_{1} |\sigma'\rangle_{2}$ and the pseudo-fermion 
operators $f_{1\sigma}^{\dagger}=|\sigma\rangle_{1} |0\rangle_{2}$, 
$f_{2\sigma}^{\dagger}=|0\rangle_{1} |\sigma\rangle_{2}$. Then, 
Eqs.~(\ref{sf3})-(\ref{comp3}) can be replaced as
\bn
c_{1\sigma}&=& e^{\dagger} f_{1\sigma} + \sum_{\sigma'} f_{2\sigma'}^{\dagger} d_{\sigma 
\sigma'}^{\pdag}, \\
c_{2\sigma}&=& e^{\dagger} f_{2\sigma} + \sum_{\sigma'} f_{1\sigma'}^{\dagger} 
d_{\sigma' \sigma}^{\pdag},
\en
\bq
e^{\dagger} e+ \sum_{\sigma} (f_{1\sigma}^{\dagger} f_{1 \sigma}^{\pdag} + 
f_{2\sigma}^{\dagger} f_{2\sigma}^{\pdag}) + \sum_{\sigma \sigma'} d_{\sigma 
\sigma'}^{\dagger} d_{\sigma \sigma'}^{\pdag}=1.
\eq
And obviously the explicit (anti)communicators of these auxiliary particles are:
\[
ee^{\dagger}=1, \quad d_{\sigma_1 \sigma_2}d_{\sigma_1' \sigma_2'}^{\dagger} = 
\delta_{\sigma_1 \sigma_1'} \delta_{\sigma_2 \sigma_2'}, \quad f_{i\sigma}^{\pdag} 
f_{j\sigma'}^{\dagger}=\delta_{ij} \delta_{\sigma \sigma'},
\]
\bq
ed_{\sigma \sigma'}^{\dagger} = e f_{j\sigma}^{\dagger} = f_{j\sigma}^{\pdag}e^{\dagger} 
= f_{j\sigma}^{\pdag} d_{\sigma' \sigma^{\prime\prime} }^{\dagger} = d_{\sigma 
\sigma'}e^{\dagger} = d_{\sigma' \sigma^{\prime\prime} } f_{j\sigma}^{\dagger}=0. 
\label{quan3}
\eq 
The density matrix elements are expressed as $\hat \rho_{00}=|0\rangle_{1} |0 
\rangle_{2} {}_{2}\langle 0| {}_{1}\langle 0|=e^{\dagger} e$, $\hat 
\rho_{11\sigma}=|\sigma\rangle_{1} |0 \rangle_{2} {}_{2}\langle 0| {}_{1}\langle \sigma| 
=f_{1\sigma}^{\dagger} f_{1\sigma}^{\pdag}$, $\hat \rho_{22\sigma}=|0\rangle_{1} |\sigma 
\rangle_{2} {}_{2}\langle \sigma| {}_{1}\langle 0|=f_{2\sigma}^{\dagger} 
f_{2\sigma}^{\pdag}$, $\hat \rho_{dd\sigma \sigma'}=|\sigma\rangle_{1} |\sigma' 
\rangle_{2} {}_{2}\langle \sigma'| {}_{1}\langle \sigma|=d_{\sigma \sigma'}^{\dagger} 
d_{\sigma \sigma'}^{\pdag}$, and $\hat \rho_{12\sigma}=|0\rangle_{1} |\sigma \rangle_{2} 
{}_{2}\langle 0| {}_{1}\langle \sigma|= f_{2\sigma}^{\dagger} f_{1\sigma}^{\pdag}$. In 
terms of these slave particles operators, the Hamiltonian for the CQD can be rewritten 
as
\bn
H &=& \sum_{\eta, k, \sigma}\epsilon _{\eta k\sigma} 
c_{\eta k\sigma }^{\dagger }c_{\eta k\sigma }^{\pdag} + \epsilon_{1} \sum_{\sigma} f_{1 
\sigma }^{\dagger } f_{1 \sigma }^{\pdag} + \epsilon_{2} \sum_{\sigma} f_{2 \sigma 
}^{\dagger } f_{2 \sigma }^{\pdag} \cr
&& + t\sum_{\sigma} (f_{1 \sigma}^{\dagger} f_{2\sigma}^{\pdag} + f_{2 \sigma}^{\dagger} 
f_{1\sigma}^{\pdag}) + (2\epsilon_{d} +U) \sum_{\sigma, \sigma'} d_{\sigma 
\sigma'}^{\dagger} d_{\sigma \sigma'}^{\pdag} \cr
&& + \sum_{k, \sigma} [V_{L \sigma} c_{L k\sigma }^{\dagger } (e^{\dagger} f_{1 \sigma} 
+ \sum_{\sigma'} f_{2\sigma'}^{\dagger} d_{\sigma \sigma'}^{\pdag}) +{\rm {H.c.}}] \cr 
&& + \sum_{k, \sigma} [V_{R \sigma} c_{R k\sigma }^{\dagger } (e^{\dagger} f_{2 \sigma} 
+ \sum_{\sigma'} f_{1\sigma'}^{\dagger} d_{\sigma' \sigma}^{\pdag}) +{\rm {H.c.}}].
\label{hamiltonian4}
\en

\subsection{The quantum rate equations for the CQD}

Define the retarded (advanced) and lesser (greater) Green's functions (GFs) for the CQD 
$G_{ij \sigma }^{r(a),<(>)}(t,t')\equiv \langle \langle c_{i\sigma}^{\pdag} (t) | 
c_{j\sigma}^{\dagger}(t') \rangle\rangle ^{r(a),<(>)}$ as usual. Considering 
Eqs.~(\ref{sf3}) and (\ref{sf4}), these GFs can be expressed in terms of the slave 
particles: $G_{ij \sigma}=G_{e ij \sigma}+ \sum_{\sigma'\sigma^{\prime\prime} } G_{d 
\bar{i} \bar{j} \sigma \sigma'\sigma^{\prime\prime} } $ [$\bar{i}=2(1)$ if $i=1(2)$] 
with $G_{e ij \sigma}\equiv \langle \langle e^{\dagger}(t) f_{i\sigma}^{\pdag} (t)| 
f_{j\sigma}^{\dagger}(t') e(t')\rangle \rangle$ and $G_{d 11 \sigma 
\sigma'\sigma^{\prime\prime} }= \langle \langle f_{1 \sigma'}^{\dagger}(t) d_{\sigma' 
\sigma}^{\pdag} (t) | d_{\sigma^{\prime\prime}  \sigma}^{\dagger}(t') f_{1 
\sigma^{\prime\prime} }^{\pdag} (t')\rangle \rangle$, $G_{d 22 \sigma 
\sigma'\sigma^{\prime\prime} }= \langle \langle f_{2 \sigma'}^{\dagger}(t) d_{\sigma 
\sigma'}^{\pdag} (t) | d_{\sigma \sigma^{\prime\prime} }^{\dagger}(t') f_{2 
\sigma^{\prime\prime} }^{\pdag} (t')\rangle \rangle$. In the following derivation, we 
will use the nondiagonal doubly-occupied-related GFs, for example, $G_{d 21 \sigma 
\sigma'\sigma^{\prime\prime} }= \langle \langle f_{2 \sigma^{\prime\prime}}^{\dagger}(t) 
d_{\sigma' \sigma^{\prime\prime}}^{\pdag} (t) | d_{\sigma' \sigma}^{\dagger}(t') f_{1 
\sigma }^{\pdag} (t')\rangle \rangle$ and $G_{d 21 \sigma \sigma'\sigma^{\prime\prime} 
}'= \langle \langle f_{2 \sigma}^{\dagger}(t) d_{\sigma \sigma'}^{\pdag} (t) | 
d_{\sigma^{\prime\prime} \sigma' }^{\dagger}(t') f_{1 \sigma^{\prime\prime} }^{\pdag} 
(t')\rangle \rangle$. Under the weak coupling assumption and small bare detuning 
$\delta$, the decoupled GFs of the CQD can be defined in terms of spectrum expressions, 
in the Fourier space, as
\bn
G_{e ii\sigma}^{<0}(\omega)&=& 2\pi i \rho_{ii\sigma } \delta (\omega- \epsilon_{d}), 
\cr
G_{d 11 \sigma \sigma' \sigma^{\prime\prime} }^{<0}(\omega) &=& \delta_{\sigma' 
\sigma^{\prime\prime} } 2\pi i \rho_{dd \sigma' \sigma} \delta [\omega - (\epsilon_{d} 
+U)], \cr
G_{d 22 \sigma \sigma' \sigma^{\prime\prime} }^{<0}(\omega) &=& \delta_{\sigma' 
\sigma^{\prime\prime} } 2\pi i \rho_{dd \sigma \sigma'} \delta [\omega - (\epsilon_{d} 
+U)], \cr
G_{e ii\sigma }^{>0}(\omega) &=& -2\pi i \rho_{00} \delta (\omega - \epsilon_{d}), \cr
G_{d ii\sigma \sigma' \sigma^{\prime\prime} }^{>0}(\omega) &=& - \delta_{\sigma' 
\sigma^{\prime\prime} } 2\pi i \rho_{ii \sigma'} \delta [\omega-(\epsilon_{d}+U)], \cr
&& \label{gg1}
\en
\bn
G_{d ij\sigma \sigma' \sigma^{\prime\prime} }^{(\prime)<0}(\omega) &=& 0, \quad G_{e 
ij\sigma }^{>0}(\omega) = 0. \cr
G_{e ij\sigma }^{<0}(\omega) &=& 2\pi i \rho_{ij\sigma } \delta(\omega - \epsilon_{d}), 
\cr
G_{d ij\sigma \sigma' \sigma^{\prime\prime} }^{(\prime)>0}(\omega) &=& - \delta_{\sigma 
\sigma^{\prime\prime}} 2\pi i \rho_{ji \sigma} \delta[\omega - (\epsilon_{d}+U)]. 
\label{gg2} 
\en
In order to get the quantum rate equations, we use exactly the same procedure as in the 
previous section, evaluating the statistical expectations of the rate of time change of 
the density matrix elements $\rho_{ij}$. After tedious but straightforward calculations, 
eventually we obtain in the wide band limit 
\begin{widetext}
\begin{subequations}
\bq
\dot{\rho}_{00}= -\frac{i}{2\pi} \int d\omega \sum_{\sigma} \Big \{ \Gamma_{L \sigma} 
f_{L}(\omega)G_{e 11\sigma}^{>}(\omega)
+ \Gamma_{L \sigma} [1-f_{L}(\omega)] G_{e 11\sigma}^{<}(\omega) + \Gamma_{R \sigma} 
f_{R}(\omega)G_{e 22\sigma}^{>}(\omega)
+ \Gamma_{R \sigma} [1-f_{R}(\omega)] G_{e 22\sigma}^{<}(\omega) \Big \}, 
\eq
\bn
\dot{\rho}_{11 \sigma}&=& \frac{i}{2\pi} \int d\omega \Big \{ \Gamma_{L \sigma} 
f_{L}(\omega) G_{e 11\sigma}^{>}(\omega)
+ \Gamma_{L \sigma} [1-f_{L}(\omega)] G_{e 11\sigma}^{<}(\omega) 
- \sum_{\sigma', \sigma^{\prime\prime} } \Gamma_{R \sigma'} f_{R}(\omega) G_{d 11 
\sigma' \sigma \sigma^{\prime\prime} }^{>}(\omega) \cr 
&& - \sum_{\sigma', \sigma^{\prime\prime} } \Gamma_{R \sigma'} [1-f_{R}(\omega)] G_{d 11 
\sigma' \sigma \sigma^{\prime\prime} }^{<}(\omega) \Big \} + it (\rho_{12\sigma} - 
\rho_{21\sigma}), \\
\dot{\rho}_{22 \sigma}&=& \frac{i}{2\pi} \int d\omega \Big \{ \Gamma_{R \sigma} 
f_{R}(\omega) G_{e 22\sigma}^{>}(\omega)
+ \Gamma_{R \sigma} [1-f_{R}(\omega)] G_{e 22\sigma}^{<}(\omega) 
- \sum_{\sigma', \sigma^{\prime\prime} } \Gamma_{L \sigma'} f_{L}(\omega) G_{d 22 
\sigma' \sigma \sigma^{\prime\prime} }^{>}(\omega) \cr 
&&  - \sum_{\sigma', \sigma^{\prime\prime} } \Gamma_{L \sigma'} [1-f_{L}(\omega)] G_{d 
22 \sigma' \sigma \sigma^{\prime\prime} }^{<}(\omega) \Big \} + it (\rho_{21\sigma} - 
\rho_{12\sigma}),
\en
\bn
\dot{\rho}_{12\sigma }&=& i(\epsilon_2-\epsilon_1)\rho_{12\sigma} + \frac{i}{4\pi} \int 
d\omega \Big \{ \sum_{\eta} \{ \Gamma_{\eta \sigma} f_{\eta}(\omega) G_{e 12\sigma 
}^{>}(\omega)
+ \Gamma_{\eta \sigma} [1-f_{\eta}(\omega)] G_{e 12\sigma }^{<}(\omega) \} \cr
&& - \sum_{\sigma', \sigma^{\prime\prime} } \{ \Gamma_{L \sigma'} f_{L}(\omega) G_{d21 
\sigma \sigma' \sigma^{\prime\prime} }^{ >}(\omega) + \Gamma_{L \sigma'} 
[1-f_{L}(\omega)] G_{d21 \sigma \sigma' \sigma^{\prime\prime} }^{ <}(\omega) \} \cr
&&  - \sum_{\sigma', \sigma^{\prime\prime} } \{ \Gamma_{R \sigma'} f_{R}(\omega) G_{d21 
\sigma \sigma' \sigma^{\prime\prime} }^{\prime >}(\omega) + \Gamma_{R \sigma'} 
[1-f_{R}(\omega)] G_{d21 \sigma \sigma' \sigma^{\prime\prime} }^{\prime <}(\omega) \} 
\Big \} + it(\rho_{11 \sigma}-\rho_{22\sigma}),
\en
\bq
\dot{\rho}_{dd \sigma \sigma}= \frac{i}{2\pi} \int d\omega \sum_{\sigma'} \Big \{  
\Gamma_{L \sigma} f_{L}(\omega) G_{d 22\sigma \sigma \sigma'}^{>}(\omega) + \Gamma_{L 
\sigma} [1-f_{L}(\omega)] G_{d 22 \sigma \sigma \sigma'}^{<}+ \Gamma_{R \sigma} 
f_{R}(\omega) G_{d 11\sigma \sigma \sigma'}^{>}(\omega) + \Gamma_{R \sigma} 
[1-f_{R}(\omega)] G_{d 11 \sigma \sigma \sigma'}^{<} \Big \},
\eq
\bq
\dot{\rho}_{dd \sigma \bar{\sigma}}= \frac{i}{2\pi} \int d\omega \sum_{\sigma'}\Big \{ 
\Gamma_{L \sigma} f_{L}(\omega) G_{d 22\sigma \bar{\sigma} \sigma'}^{>}(\omega) + 
\Gamma_{L \sigma} [1-f_{L}(\omega)] G_{d 22 \sigma \bar{\sigma} \sigma'}^{<} + \Gamma_{R 
\bar{\sigma}} f_{R}(\omega) G_{d 11 \bar{\sigma} \sigma \sigma'}^{>}(\omega) + \Gamma_{R 
\bar{\sigma}} [1-f_{R}(\omega)] G_{d 11 \bar{\sigma} \sigma \sigma'}^{<} \Big \}. 
\eq
\end{subequations}
\end{widetext}
Substituting these correlation GFs with their decoupled formulations Eqs.~(\ref{gg1}) 
and (\ref{gg2}), the quantum rate equations can be obtained
\begin{subequations}
\label{rateqCQD}
\bn
\dot{\rho}_{00}&=& \sum_{\sigma} [ \Gamma_{L \sigma}^{-} \rho_{11\sigma} + \Gamma_{R 
\sigma}^{-} \rho_{22\sigma} - (\Gamma_{L \sigma}^{+} + \Gamma_{R \sigma}^{+}) 
\rho_{00}], \cr
&& \label{rc0} \\
\dot{\rho}_{11\sigma}&=& \Gamma_{L\sigma}^{+} \rho_{00} + \sum_{\sigma'} 
\widetilde{\Gamma}_{R \sigma'}^{-} \rho_{dd \sigma\sigma'} - \Gamma_{L\sigma}^{-} 
\rho_{11\sigma} \cr
&& - \sum_{\sigma'} \widetilde{\Gamma}_{R \sigma'}^{+} \rho_{11\sigma} - 2t \Im 
\rho_{12\sigma}, \label{rc1} \\
\dot{\rho}_{22\sigma}&=& \Gamma_{R\sigma}^{+} \rho_{00} + \sum_{\sigma'} 
\widetilde{\Gamma}_{L \sigma'}^{-} \rho_{dd \sigma'\sigma}- \Gamma_{R\sigma}^{-} 
\rho_{22\sigma} \cr
&& - \sum_{\sigma'}\widetilde{\Gamma}_{L\sigma'}^{+} \rho_{22\sigma} + 2t \Im 
\rho_{12\sigma}, \label{rc2} \\
\dot{\rho}_{12\sigma}&=& i(\epsilon_2-\epsilon_1) \rho_{12\sigma} +it (\rho_{11\sigma}- 
\rho_{22\sigma}) \cr
&& - \frac{1}{2} [\Gamma_{L\sigma}^{-} + \Gamma_{R\sigma}^{-} + \sum_{\eta, \sigma'} 
\widetilde{\Gamma}_{\eta \sigma'}^{+}] \rho_{12\sigma}, \label{rc3}
\en
\bq
\dot{\rho}_{dd\sigma \sigma}= \widetilde{\Gamma}_{R\sigma}^{+} \rho_{11\sigma} + 
\widetilde{\Gamma}_{L\sigma}^{+} \rho_{22\sigma} - (\widetilde{\Gamma}_{L\sigma}^{-} + 
\widetilde{\Gamma}_{R\sigma}^{-}) \rho_{dd \sigma \sigma}, \label{rc4}
\eq
\bq
\dot{\rho}_{dd\sigma \bar{\sigma}}= \widetilde{\Gamma}_{R \bar{\sigma}}^{+} 
\rho_{11\sigma} + \widetilde{\Gamma}_{L\sigma}^{+} \rho_{22 \bar{\sigma}} - 
(\widetilde{\Gamma}_{L\sigma}^{-} + \widetilde{\Gamma}_{R \bar{\sigma}}^{-}) \rho_{dd 
\sigma \bar{\sigma}}, \label{rc5}
\eq
\end{subequations}
and along with $\rho_{00}+ \sum_{\sigma} (\rho_{11\sigma}+ \rho_{22\sigma}) + 
\sum_{\sigma, \sigma'} \rho_{dd\sigma \sigma'}=1$, in whcih $\Gamma_{\eta\sigma}^{\pm}= 
\Gamma_{\eta \sigma} f_{\eta}^{\pm}(\epsilon_{d})$ and $\widetilde{\Gamma}_{\eta 
\sigma}^{\pm}= \Gamma_{\eta \sigma} f_{\eta}^{\pm}(\epsilon_{d}+U)$ have the similar 
prescriptions as in the SQD. In addition, the classical parts of the diagonal elements 
equations have the similar interpretations. For example, Eq.~(\ref{rc1}) for the rate of 
change of the number of the spin-$\sigma$ electrons in the first QD $\rho_{11 \sigma}$ 
is contributed, noting the fact that the first (second) QD do not directly connect to 
the right (left) lead, from four single-particle tunneling processes: 1) tunneling into 
the QD with spin-$\sigma$ electrons $\Gamma_{L\sigma}^{+}$ from the left lead if the QD 
is initially in the empty state $\rho_{00}$; 2) tunneling out from the QD with 
spin-$\sigma'$ electrons $\widetilde {\Gamma}_{R \sigma'}^{-}$ into the right lead if 
the QD is initially in the doubly occupied state $\rho_{dd \sigma \sigma'}$; 3) 
tunneling into the QD with spin-$\sigma'$ electrons $\widetilde {\Gamma}_{R 
\sigma'}^{+}$ from the right lead; and 4) tunneling out from the QD with spin-$\sigma$ 
electrons $\Gamma_{L\sigma}^{-}$ into the left lead, when the QD is initially just in 
this state $\rho_{11 \sigma}$. Tunneling events 1) and 2) increase $\rho_{11 \sigma}$, 
but events 3) and 4) decrease this probability. The final term in Eq.~(\ref{rc1}) is 
responsible for coherent effects. The equation (\ref{rc3}) for the nondiagonal matrix 
element $\rho_{12 \sigma}$ indicates that the role of the leads is to provide damping of 
the quantum superposition.\cite{Gurvitz} It is also worth noting that the present 
proposed quantum rate equations are reliable for a wide range of temperature and 
external bias voltage, where the three major approximations we use are valid.    

The electric current $I_{L}$ flowing from the lead $L$ to the QD can be calculated as:
\bn
I_{L}&=& ie\int \frac{d\omega}{2\pi} \sum_{\sigma} \{ \Gamma_{\eta \sigma} 
f_{\eta}(\omega) [G_{e11 \sigma}^{>}(\omega) \cr
&& + \sum_{\sigma', \sigma^{\prime\prime} } G_{d 22 \sigma \sigma' \sigma^{\prime\prime} 
}^{>}(\omega)] + \Gamma_{\eta \sigma} [1-f_{\eta}(\omega)] [G_{e11 \sigma}^{<}(\omega) 
\cr
&& + \sum_{\sigma', \sigma^{\prime\prime} } G_{d22 \sigma \sigma' \sigma^{\prime\prime} 
}^{<}] \}.  
\en
Under the weak coupling approximation, it becomes
\begin{subequations}
\bn
I_{L}/e&=& \sum_{\sigma} [ \widetilde{\Gamma}_{L \sigma}^{-} (\rho_{dd \sigma \sigma} + 
\rho_{dd \sigma \bar{\sigma}}) + \Gamma_{L \sigma}^{-} \rho_{11\sigma } \cr
&& - \widetilde{\Gamma}_{L \sigma}^{+} (\rho_{22 \sigma}+ \rho_{22 \bar{\sigma}}) - 
\Gamma_{L \sigma}^{+} \rho_{00}]. \label{iiil}
\en 
Similarly, for the current flowing from the lead $R$ we have
\bn
I_{R}/e&=& \sum_{\sigma} [ \widetilde{\Gamma}_{R \sigma}^{-} (\rho_{dd \sigma \sigma} + 
\rho_{dd \bar{\sigma} \sigma}) + \Gamma_{R \sigma}^{-} \rho_{22\sigma } \cr
&& - \widetilde{\Gamma}_{R \sigma}^{+} (\rho_{11 \sigma}+ \rho_{11 \bar{\sigma}}) - 
\Gamma_{R \sigma}^{+} \rho_{00}]. \label{iiir}
\en
\end{subequations}
It is easy to prove that, in stationary condition, the current conservation is fulfilled 
$I_{L}=-I_{R}$. 

\subsection{Discussions}

In order to simplify the analysis, we only consider spin independent tunneling processes 
in the following discussions. Two special cases, no doubly occupied state and no empty 
state, are studied. First we assume the interdot Coulomb interaction $U$ is infinite, 
whereas only one electron can be found inside the system, so $\rho_{dd\sigma \sigma'}=0$ 
and $\widetilde{\Gamma}_{\eta \sigma}^{+}\simeq 0$. The quantum rate equations 
(\ref{rc1})-(\ref{rc5}) simplify to
\begin{subequations}
\bn
\dot{\rho}_{11}&=& \Gamma_{L}^{+} \rho_{00} - \Gamma_{L}^{-} \rho_{11} - 2t\Im 
\rho_{12}, \\
\dot{\rho}_{22}&=& \Gamma_{R}^{+} \rho_{00} - \Gamma_{R}^{-} \rho_{22} + 2t\Im  
\rho_{12},
\en
\bq
\dot{\rho}_{12}= i(\epsilon_2-\epsilon_1) \rho_{12} + it (\rho_{11} - \rho_{22}) - 
\frac{1}{2}(\Gamma_{L}^{-} + \Gamma_{R}^{-}) \rho_{12},
\eq
\end{subequations}
with $\rho_{00}+2 \rho_{11}+ 2\rho_{22}=1$. The steady solutions are
\begin{subequations}
\bn
\hspace{-1cm}\rho_{11}&=& [\Gamma_{L}^{+} \Gamma_{R}^{-} + t^2 (\Gamma_{L}^{+}+ 
\Gamma_{R}^{+}) (\Gamma_{L}^{-} + \Gamma_{R}^{-})/\Lambda]/\Delta,\\   
\hspace{-1cm}\rho_{22}&=& [\Gamma_{L}^{-} \Gamma_{R}^{+} + t^2 (\Gamma_{L}^{+}+ 
\Gamma_{R}^{+}) (\Gamma_{L}^{-} + \Gamma_{R}^{-})/\Lambda]/\Delta,
\en
\bq
\rho_{12}= t(\Gamma_{L}^{+} \Gamma_{R}^{-} - \Gamma_{L}^{-} 
\Gamma_{R}^{+})[\epsilon_1-\epsilon_2 + i \frac{1}{2} (\Gamma_{L}^{-}+ 
\Gamma_{R}^{-})]/\Delta\Lambda,
\eq
\bn 
\hspace{-1cm}\Delta&=& \Gamma_{L}^{-} \Gamma_{R}^{-} + 2\Gamma_{L}^{-} \Gamma_{R}^{+} + 
2\Gamma_{L}^{+} \Gamma_{R}^{-} \cr
\hspace{-1cm}&& + t^2 (\Gamma_{L}^{-}+ \Gamma_{R}^{-}+ 4\Gamma_{L}^{+} +  
4\Gamma_{R}^{+})(\Gamma_{L}^{-} + \Gamma_{R}^{-}) /\Lambda,   
\en
\end{subequations}
in which $\Lambda=(\epsilon_2-\epsilon_1)^2 + (\Gamma_{L}^{-}+ \Gamma_{R}^{-})^2/4$. The 
steady current is given by $I_{L}/e=2(\Gamma_{L}^{-} \rho_{11}- \Gamma_{L}^{+} 
\rho_{00})$.

It is interesting to compare our results in this situation with those of Gurvitz and 
Prager\cite{Gurvitz} for the case of large bias voltage between the two leads. For 
example, the large bias voltage determines $\Gamma_{L}^{-}=0$, $\Gamma_{R}^{+}=0$ and 
$\Gamma_{L}^{+}=\Gamma_{L}$, $\Gamma_{R}^{-}=\Gamma_{R}$ at $eV\gg T$. Therefore, the dc 
current becomes
\bq
I_{L}/e=-\frac{t^2 \Gamma_{R}} {t^2 (2+ \Gamma_{R}/2\Gamma_{L})+ (\Gamma_{R})^2/4+ 
(\epsilon_2-\epsilon_1)^2},
\eq  
which coincides with the result obtained by Gurvitz and Prager.\cite{Gurvitz} It is 
quite obvious that the finite temperature plays a crucial role in the ``coherence" 
tunneling. The previous formulations for large bias voltages, however, can not provide 
any information about the temperature effects. This is the central improvement of the 
present approach for the coupled quantum systems.

Secondly, we consider the deep level situation where the bare levels $\epsilon_1$ and 
$\epsilon_2$ are far below the Fermi level but $\epsilon_{d}+U$ is just above the Fermi 
level in equilibrium. In this case the CQD is always occupied and $\rho_{00}=0$, 
$\Gamma_{\eta \sigma}^{-}\simeq 0$. Therefore, we have
\begin{subequations}
\bn
\dot{\rho}_{11}&=& 2\widetilde{\Gamma}_{R}^{-} \rho_{dd} - 2\widetilde{\Gamma}_{R}^{+} 
\rho_{11} - 2t\Im \rho_{12}, \\
\dot{\rho}_{22}&=& 2\widetilde{\Gamma}_{L}^{-} \rho_{dd} - 2\widetilde{\Gamma}_{L}^{+} 
\rho_{22} + 2t\Im \rho_{12}, \\
\dot{\rho}_{dd}&=& \widetilde{\Gamma}_{R}^{+} \rho_{11} + \widetilde{\Gamma}_{L}^{+} 
\rho_{22} - (\widetilde{\Gamma}_{L}^{-} + \widetilde{\Gamma}_{R}^{-}) \rho_{dd},
\en
\bq
\dot{\rho}_{12}= i(\epsilon_2-\epsilon_1) \rho_{12} + it(\rho_{11} - \rho_{22}) -  
(\widetilde{\Gamma}_{L}^{+} + \widetilde{\Gamma}_{R}^{+}) \rho_{12},
\eq
\end{subequations}
with $2\rho_{11}+ 2\rho_{22}+ 4\rho_{dd}=1$. After solving the set of equations in the 
steady state, we obtain
\begin{subequations}
\bn
\rho_{11}&=& [\widetilde{\Gamma}_{L}^{+} \widetilde{\Gamma}_{R}^{-} + t^2 
(\widetilde{\Gamma}_{L}^{+}+ \widetilde{\Gamma}_{R}^{+}) (\widetilde{\Gamma}_{L}^{-} + 
\widetilde{\Gamma}_{R}^{-})/\Lambda]/\Delta,\\   
\rho_{22}&=& [\widetilde{\Gamma}_{L}^{-} \widetilde{\Gamma}_{R}^{+} + t^2 
(\widetilde{\Gamma}_{L}^{+}+ \widetilde{\Gamma}_{R}^{+}) (\widetilde{\Gamma}_{L}^{-} + 
\widetilde{\Gamma}_{R}^{-})/\Lambda]/\Delta,\\
\rho_{12}&=& t(\widetilde{\Gamma}_{L}^{+} \widetilde{\Gamma}_{R}^{-} - 
\widetilde{\Gamma}_{L}^{-} \widetilde{\Gamma}_{R}^{+})[\epsilon_1-\epsilon_2 + i 
(\widetilde{\Gamma}_{L}^{+}+ \widetilde{\Gamma}_{R}^{+})]/\Delta\Lambda, \cr
&& \\ 
\Delta&=& 2\widetilde{\Gamma}_{L}^{+} \widetilde{\Gamma}_{R}^{-} + 
2\widetilde{\Gamma}_{L}^{-} \widetilde{\Gamma}_{R}^{+} + 4\widetilde{\Gamma}_{L}^{+} 
\widetilde{\Gamma}_{R}^{+} \cr
&& + 4t^2 (\widetilde{\Gamma}_{L}^{-}+ \widetilde{\Gamma}_{R}^{-}+ 
\widetilde{\Gamma}_{L}^{+}+ \widetilde{\Gamma}_{R}^{+} ) (\widetilde{\Gamma}_{L}^{+} + 
\widetilde{\Gamma}_{R}^{+}) /\Lambda,   
\en
\end{subequations}
in which $\Lambda=(\epsilon_2-\epsilon_1)^2 + (\widetilde{\Gamma}_{L}^{+}+ 
\widetilde{\Gamma}_{R}^{+})^2$. The dc current is $I_{L}/e=4(\widetilde{\Gamma}_{L}^{-} 
\rho_{dd} - \widetilde{\Gamma}_{L}^{+} \rho_{22})$.

It is also interesting to consider the situation of large bias voltage in the strong 
interdot Coulomb repulsion $U$, whereas the Fermi level of the right lead $\mu_{R}$ lies 
far below $\epsilon_d+U$, but far above the resonance level $\epsilon_d$ to satisfy the 
requirement of deep level, meanwhile the Fermi level of the left lead $\mu_{L}$ is far 
above $\epsilon_{d}+U$, so that $\widetilde{\Gamma}_{L}^{-}=0$, 
$\widetilde{\Gamma}_{R}^{+}=0$ and $\widetilde{\Gamma}_{L}^{+}=\Gamma_{L}$, 
$\widetilde{\Gamma}_{R}^{-}=\Gamma_{R}$. Finally we obtain
\bq
I_{L}/e=-\frac{2t^2 \widetilde{\Gamma}_{L}} {2t^2 (1+ \widetilde{\Gamma}_{L}/ 
\widetilde{\Gamma}_{R}) + 2(\epsilon_2- \epsilon_{1})^2 + (\widetilde{\Gamma}_{L})^2}.
\eq  

\subsection{Numerical results}

In this subsection, we perform numerical calculations for the tunneling transport 
through the CQD, by using the quantum rate equations (\ref{rateqCQD}), in the stationary 
condition. We symmetrically add the bias voltage again between the source and drain 
$\mu_{L}=-\mu_{R}=eV/2$.   

First we consider the spin independent transport. Fig.~5 demonstrates the nonequilibrium 
occupation numbers in the first and the second QDs, calculated from the obtained 
expectation values of density-matrix elements 
$n_{1\sigma}=\rho_{11\sigma}+\sum_{\sigma'}\rho_{dd\sigma \sigma'}$ and 
$n_{2\sigma}=\rho_{22\sigma}+\sum_{\sigma'}\rho_{dd\sigma' \sigma}$, and the 
corresponding current versus the discrete level for the hopping $t=1.0$ between the two 
QDs at a small bias $V=1.0$ and a large bias $V=10.0$, respectively. We find a similar 
characteristic as in the SQD (Fig.~2): 1) The nonzero bias weakens the Coulomb blockade 
effect; 2) $n_{1\sigma}$ and $n_{2\sigma}$ have fractional steps and 3) $n_{1\sigma}\neq 
n_{2\sigma}$ in nonequilibrium; 4) The conductance has two peaks at the resonant points, 
while the current has three steps in the strong nonequilibrium regime. Here we observe a 
higher peak magnitude and a higher step value at the deep level regime than those at the 
no doubly occupied level regime, and the maximum step value located at the middle 
``window" of the bare level. More interestingly, a opposite behavior has been found when 
the hopping $t$ between two QDs decreases, as shown in Fig.~6(c), in which we plot the 
corresponding results for a small hopping $t=0.5$. Generally, one may expect that 
increasing the hopping $t$ can reduce the difference between two QDs, and very strong 
hopping can finally give rise to the formation of covalence. In other words, the 
difference between $n_{1\sigma}$ and $n_{2\sigma}$ should raising with decreasing the 
hopping $t$. This is the case as shown in Figs.~6(a) and (b), where the occupation 
numbers are displayed for the smaller hopping $t=0.5$ in comparison with the results of 
the hopping $t=1.0$ in Figs.~5(a) and (b). One can note that the occupation number in 
the second QD even experiences a descendance in the middle ``window" of the bare level 
for the case of $t=0.5$, which expresses an opposite behavior in the case of $t=1.0$. 
This is the reason why current-voltage characteristics are different in the two cases.

The effect of the hopping on the tunneling is more clearly illustrated in Fig.~7, where 
we plot the occupation numbers, the current, and the differential conductance as a 
function of the bias for different hoppings $t$ in the no doubly occupied level 
$\epsilon_d=1$ and the deep level $\epsilon_d=-5$. In both cases, we have two peaks in 
the differential conductance corresponding to the two steps in the current. More 
importantly, we find that the current declines in the second step and consequently the 
negative differential conductance (NDC) appears in the according biases when the hopping 
$t<1.0$. We can explain appearance of the NDC by variations of the occupation numbers 
with the bias, as shown in Figs.~7(a) and (d) for the hoppings $t=0.2$ (thick lines) and 
$t=1.0$ (thin lines). Considering the fact that we apply the bias symmetrically and 
$\mu_{R}=-eV/2<0$, the current flowing from the right lead is dominated for the case 
$\epsilon_{d}=1$ by the process: tunneling out from the second QD into the right lead. 
According to Eq.~(\ref{iiir}), we have $I_{R}/e \approx \sum_{\sigma} 
\Gamma_{R\sigma}n_{2\sigma}$. It is obvious from Fig.~7(a) that the rising second step 
in $n_{2\sigma}$ for the case of $t=1.0$ (thin dashed curve) indicates the rising step 
in the current, whereas the decline second step for the case of $t=0.2$ (thick dashed 
curve) implies the NDC. In the other case $\epsilon_d=-5$, the current flowing from the 
left lead is ruled by the tunneling process into the first QD from the left lead, being 
approximately $I_{L}/e=\sum_{\sigma} \Gamma_{L\sigma}[1-n_{1\uparrow}-n_{1\downarrow}]$. 
Apparently, the variations of $n_{1\sigma}$ denoted by the solid lines in Fig.~7(d) 
provide interpretations for the current-voltage characteristic in Fig.~7(e) and the NDC 
in Fig.~7(f). Therefore, it can be addressed that open of a new channel provides 
negative contribution to the current in the case of $t\lesssim 1.0$. 

An interesting question is what happens for the tunneling current and the NDC when the 
interdot Coulomb interaction $U$ weakens or strengthens. We show this in Fig.~8, where 
current vs bias is presented for various correlation parameters from $U=0$ to $\infty$ 
in the cases $t=1.0$ (thick lines) and $t=0.2$ (thin lines). For $U\rightarrow \infty$, 
the current has only one step with increasing bias in both cases of $\epsilon_{d}=1$ (a) 
and $\epsilon_{d}=-5$, because no new channel is available due to the extremely strong 
charging effect. For the finite interdot Coulomb correlation, however, the applied bias 
can overcome the Coulomb blockade effect and open a new channel for tunneling at the 
corresponding threshold value of voltage. For $t\lesssim 1.0$, this new channel induces 
a peak in the current. This peak becomes narrower with declining value of $U$, but its 
height remains unchanged if $U$ is not too small. At sufficiently small values of $U$, 
as shown in the inset of Fig.~8(a), height of the peak in current decrease, even 
vanishes finally when $U=0$. So we can claim that the interdot Coulomb interaction $U=0$ 
and $\infty$ leads to the single peak in the differential conductance, but the finite 
values result in double peaks, and even the NDC in the case of $t\lesssim 1.0$.
                                             
The temperature effect is also shown in Fig.~8 for $U=4$. Increasing temperature 
smoothes $I$-$V$ curve, but remains the NDC unchanged.

Now we study the spin dependent tunneling through the CQD connected to two ferromagnetic 
leads. Figs.~9 and 10 depict the occupation numbers in the two QDs and the current in 
both P and AP configurations for $\epsilon_{d}=1$ and $\epsilon_{d}=-5$, respectively. 
We find, besides analogous behaviors with the spin-independent tunneling, that: 1) 
$n_{1\uparrow} \neq n_{1\downarrow}$ even in both alignments; 2) $n_{2\uparrow}\simeq 
n_{2\downarrow}$ in the P configuration but $n_{2\uparrow} \neq n_{2\downarrow}$ in the 
AP configuration; 3) variations of the current flowing in different magnetic 
configurations are very sensitive to the value of the hopping between two QDs, which 
leads to 4) the negative TMR for the sufficiently small hopping $t=0.2$, as exhibited in 
Figs.~9(c) and 10(c) at certain voltages.      

\section{Conclusion}

In this paper, we have systematically derived the quantum rate equations for sequential 
tunneling from NGF, and then utilized them to investigate quantum coherent transport in 
a single QD with weak spin-flip scattering and weakly coupled QDs systems taking the 
intradot and interdot Coulomb interactions into account. In these systems, the 
superposition between different states plays a vitally important role in coherent 
tunneling processes. Directly, a kind of quantum oscillations in mesoscopic systems is 
due to this superposition effect. Now, it is believed that the master equations or the 
modified quantum rate equations, which are actually equations of motion of density 
submatrix for diagonal and nondiagonal elements, provide a successful tool to study this 
phenomenon, and even allow an analytical description. 

For this purpose, we have generalized the slave particle technique, which is developed 
previously in the single-site space and successfully applied to study the strongly 
correlated systems, into the two-site space. Based on this theoretical approach and the 
correct quantization of these artificially introduced operators, previously 
well-developed NGF for noninteracting systems has been used to construct the quantum 
rate equations when only three assumptions are made: first, the coupling between the 
central region and the leads must be weak; second, the couplings between the subsystems 
are also weak, for example, weak spin-flip scattering in SQD and weak interdot hopping 
in CQD; third, the wide band limit. The first condition makes it valid that we can keep 
only the lowest order terms in $|V|^2$ in the expansions of the equations of motion. It 
also renders the central region approximately a qusi-equilibrium isolated system, which 
facilitates the ``localized" energy spectrum expressions for the correlation GFs of 
every subsystem in the central region in combination with the second presumption. These 
approximations not withstanding, our approach is appropriate for a wide range of 
temperature and external bias voltage, and incorporation of the charging effect. 
Finally, it should be pointed out that our derivation is equivalent to the lowest-order 
gradient expansion technique.\cite{Davies}             

Employing this approach, we have studied in detail the coherent tunneling through a SQD 
and a CQD systems. We have given some analytic expressions for steady-state transport in 
two special cases: doubly-occupied prohibited state and deep level in large intra- or 
inter-dot Coulomb repulsion. Furthermore, we have compared some of our results with 
previously obtained results in the literature. For example, for resonant tunneling 
through a SQD with spin-flip scattering, our approach provides a quantum correction to 
the classical results. When there is no spin-flip scattering, our rate equations reduce 
exactly to the classical results as Glazman and Matveev,\cite{Glazman} and 
Beenakker.\cite{Beenakker} In the case of resonant tunneling through a CQD, our results 
are in perfect agreement with the previous analysis proposed by Gurvitz and 
Prager\cite{Gurvitz} under the limitation of zero temperature and large bias voltage.

In addition, we have performed numerical simulations for variations of occupation 
numbers and the current with increasing bias voltage and varying the discrete level in 
QD. We summarize the main common features as follows: 1) Occupation numbers have 
frational steps in nonequilibrium, implying that the Coulomb blockade effect is 
partially overcome by applying bias voltage; and correspondingly 2) the current-voltage 
characteristic displays two steps, giving rise to double peaks in the differential 
conductance. Especially, our calculations manifest the importance of temperature and 
bias on the spin-flip transitions in the SQD. For the CQD, a possible NDC can be reached 
if the interdot Coulomb interaction is finite and the hopping between two QDs is small 
$t\lesssim 1.0$. Besides, the TMR becomes negative in nonequilibrium for the CQD  
connected to two ferromagnetic leads if the hopping $t$ is sufficiently small.

\begin{acknowledgments} 

B. Dong and H. L. Cui are supported by the DURINT Program administered by the US Army 
Research Office. X. L. Lei is supported by Major Projects of National Natural Science 
Foundation of China, the Special Founds for Major State Basic Research Project 
(G20000683) and the Shanghai Municipal Commission of Science and Technology (03DJ14003).

\end{acknowledgments}

\appendix*

\section{Derivation of quantum rate equations for SQD}

In the Appendix, we present a detail derivation of Eqs.~(\ref{rateq2}). In the 
following, we take Eq.~(\ref{rateq2-2}) as an example. The statistical expectation of 
Eq.~(\ref{rateq1-2}) gives
\bn
\rho_{\sigma \sigma}&=& \sum_{\eta,k} [ V_{\eta \sigma} G_{e\sigma, \eta k 
\sigma}^{<}(t,t)- \bar{\sigma}V_{\eta \bar{\sigma}} G_{d\sigma, \eta k 
\bar{\sigma}}^{<}(t,t) \cr
&& -V_{\eta \sigma}^* G_{\eta k \sigma, e\sigma}^{<}(t,t) +\bar{\sigma} V_{\eta 
\bar{\sigma}}^* G_{\eta k \bar{\sigma},d\sigma}^{<}(t,t)] \cr
&& + iR_{\bar{\sigma}}^* \rho_{\sigma \bar{\sigma}}-iR_{\sigma}\rho_{\bar{\sigma} 
\sigma}. \label{rcc} 
\en 
According to Langreth's operational rules\cite{Langreth}, those hybrid correlation GFs 
are given by
\begin{subequations}
\label{hyGF}
\bn
G_{e\sigma, \eta k \sigma'}^{<}(t,t')&=& \delta_{\sigma \sigma'}\int dt_1 [G_{e\sigma 
\sigma}^{r}(t,t_1) V_{\eta \sigma'}^* g_{\eta k \sigma'}^{<}(t_1,t') \cr
&& + G_{e \sigma \sigma}^{<}(t,t_1)V_{\eta \sigma'}^* g_{\eta k \sigma'}^{a}(t_1,t')], 
\\
G_{d\sigma, \eta k \sigma'}^{<}(t,t')&=& \delta_{\sigma \bar{\sigma}'}\sigma' \int dt_1 
[G_{d\sigma \sigma}^{r}(t,t_1)V_{\eta \sigma'}^* g_{\eta k \sigma'}^{<}(t_1,t') \cr
&& + G_{d\sigma \sigma}^{<}(t,t_1)V_{\eta \sigma'}^* g_{\eta k \sigma'}^{a}(t_1,t')], \\
G_{\eta k \sigma', e\sigma}^{<}(t,t')&=& \delta_{\sigma \sigma'} \int dt_1 [g_{\eta k 
\sigma'}^{r}(t,t_1)V_{\eta \sigma'} G_{e\sigma\sigma}^{<}(t_1,t') \cr
&&+ g_{\eta k \sigma'}^{<}(t,t')V_{\eta \sigma'} G_{e\sigma\sigma}^{a}(t_1,t')], \\
G_{\eta k \sigma', d\sigma}^{<}(t,t')&=& \delta_{\sigma \bar{\sigma}'} \sigma' \int dt_1 
[g_{\eta k \sigma'}^{r}(t,t_1)V_{\eta \sigma'} G_{d\sigma\sigma}^{<}(t_1,t') \cr
&&+ g_{\eta k \sigma'}^{<}(t,t')V_{\eta \sigma'} G_{d\sigma\sigma}^{a}(t_1,t')].
\en
\end{subequations}
Substituting Eqs.~(\ref{hyGF}) into Eq.~(\ref{rcc}) and taking the Fourier 
transformation, $\rho_{\sigma \sigma}$ can be expressed as
\bn
\rho_{\sigma \sigma}&=& \frac{1}{2\pi} \int d\omega \sum_{\eta,k} |V_{\eta \sigma}|^2 
[(G_{e\sigma \sigma}^{r}(\omega) - G_{e\sigma\sigma}^{a}(\omega)) g_{\eta k 
\sigma}^{<}(\omega) \cr
&& + G_{e\sigma\sigma}^{<}(\omega) (g_{\eta k \sigma}^{a}(\omega) -g_{\eta k 
\sigma}^{r}(\omega))] \cr
&& + |V_{\eta \bar{\sigma}}|^2 [(G_{d\sigma\sigma}^{a}(\omega) - 
G_{d\sigma\sigma}^{r}(\omega)) g_{\eta k \bar{\sigma}}^{<}(\omega) \cr  
&& + G_{d\sigma\sigma}^{<}(\omega) (g_{\eta k \bar{\sigma}}^{r}(\omega)-g_{\eta k 
\bar{\sigma}}^{a}(\omega)) ] \cr
&& + iR_{\bar{\sigma}}^* \rho_{\sigma \bar{\sigma}}-iR_{\sigma}\rho_{\bar{\sigma} 
\sigma}, \label{rcc1}
\en
where $g_{\eta k \sigma}(\omega)$ are the Fourier transform of the exact GFs in the 
$\eta$th lead without the coupling to the central region. In the wide band limit, one 
has
\begin{subequations}
\label{gf-lead}
\bn
\sum_{k} |V_{\eta \sigma}|^2 g_{\eta k \sigma}^{<}(\omega)&=& i \Gamma_{\eta \sigma} 
f_{\eta}(\omega), \\
\sum_{k} |V_{\eta \sigma}|^2 g_{\eta k \sigma}^{>}(\omega)&=& -i \Gamma_{\eta \sigma} 
[1-f_{\eta}(\omega)].
\en
\end{subequations}
Substituting the GFs (\ref{gf-lead}) into Eq.~(\ref{rcc1}) and employing $G^r-G^a\equiv 
G^>-G^<$, Eq.~(\ref{rateq2-2}) can be reached. Analogously, we can derive other 
equations in (\ref{rateq2}).

\begin{figure}[htb]
\includegraphics[height=5in,width=3in]{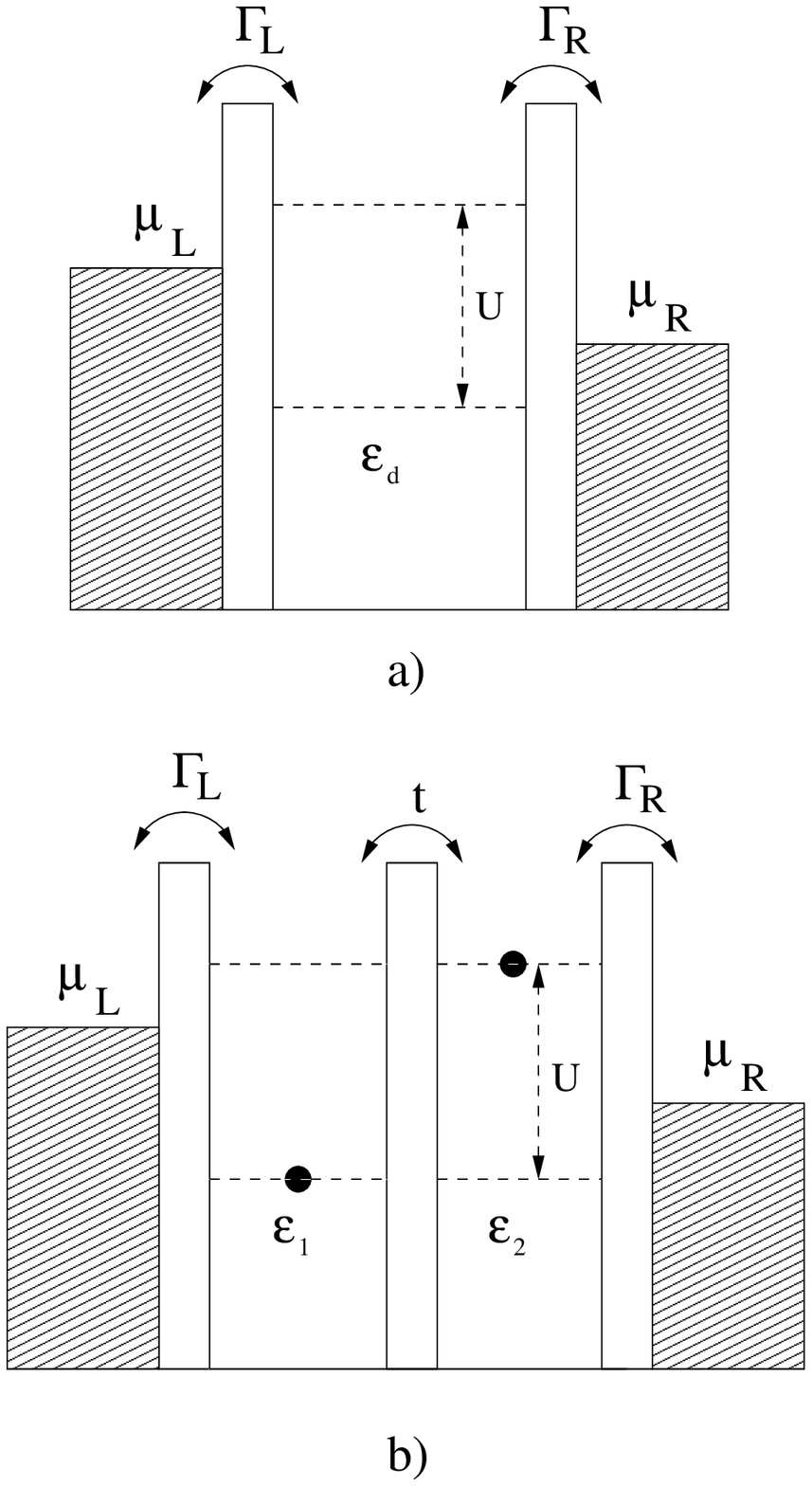}
\caption{Schematic diagrams for the resonant tunneling through (a) a single interacting 
QD and (b) a coherently coupled QDs.}
\label{fig1}
\end{figure}

\begin{figure}[htb]
\includegraphics[height=5in,width=3in]{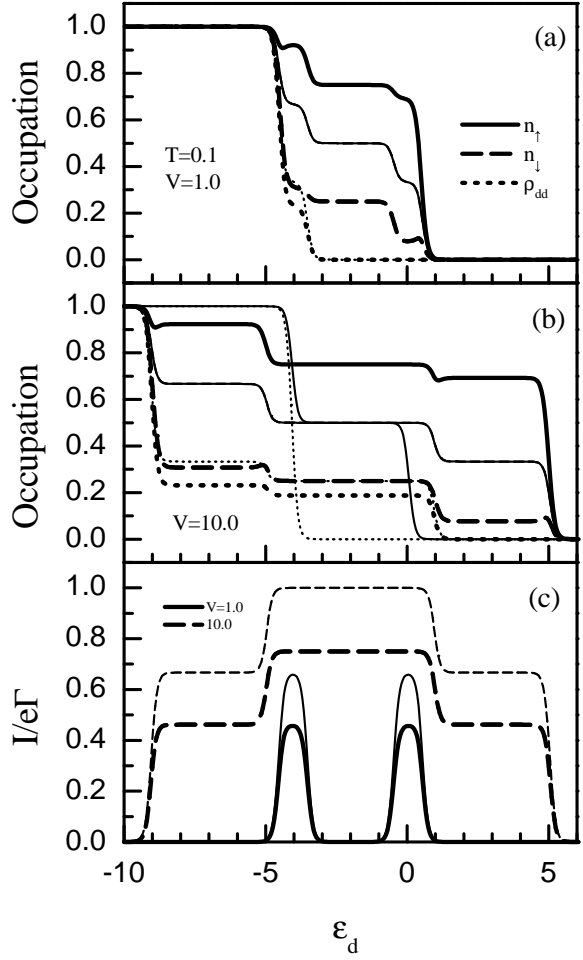}
\caption{Nonequilibrium occupation numbers $n_{\uparrow}$, $n_{\downarrow}$, and 
$\rho_{dd}$ (a,b), and tunneling current (c) vs the bare level of the SQD with no 
spin-flip scattering for both magnetization configurations. (a) is plotted at a small 
bias $V=1.0$ and (b) is at a large bias $V=10.0$. The thick lines are plotted for the AP 
configuration, and the thin curves are for the P configuration. The equilibrium 
occupation numbers are depicted by the thin lines in (b). Other parameters are: $U=4$, 
$T=0.1$, and $p=0.5$.}
\label{fig2}
\end{figure}

\begin{figure}[htb]
\includegraphics[height=5in,width=3.5in]{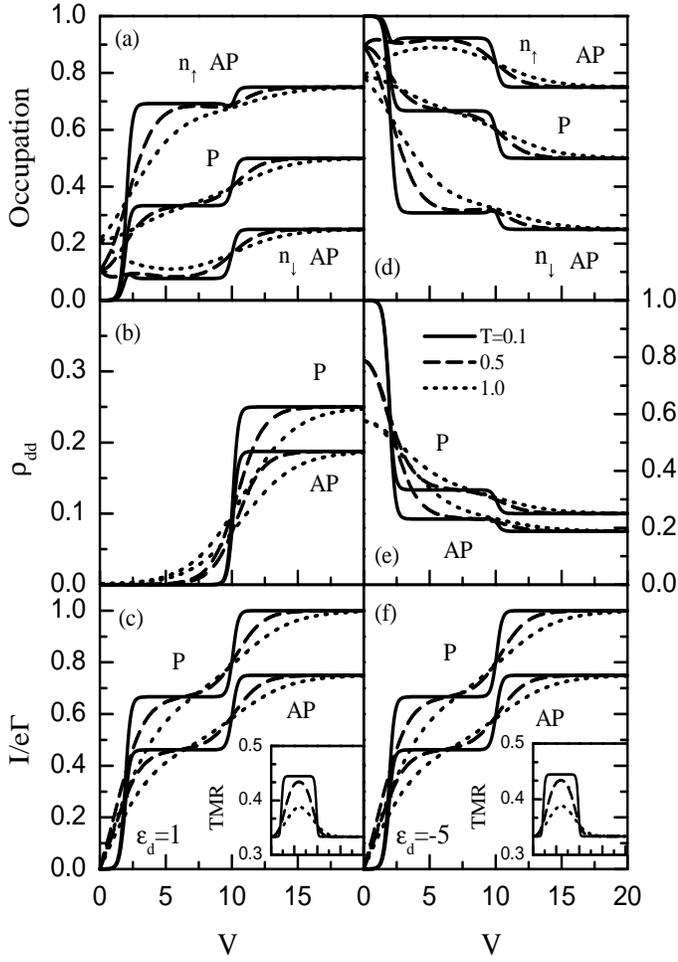}
\caption{Occupation numbers $n_{\uparrow}$, $n_{\downarrow}$ (a,d), $\rho_{dd}$ (b,e), 
and current (c,f) vs the bias voltage, calculated for no spin-flip processes and 
different temperatures $T=0.1$, $0.5$, and $1.0$. (a)-(c) are plotted for 
$\epsilon_d=1$, (d)-(f) for $\epsilon_d=-5$. The insets in (c) and (f): the 
corresponding TMR vs the bias voltage. Other parameters are as in Fig.~2.}
\label{fig3}
\end{figure}

\begin{figure}[htb]
\includegraphics[height=5in,width=3.5in]{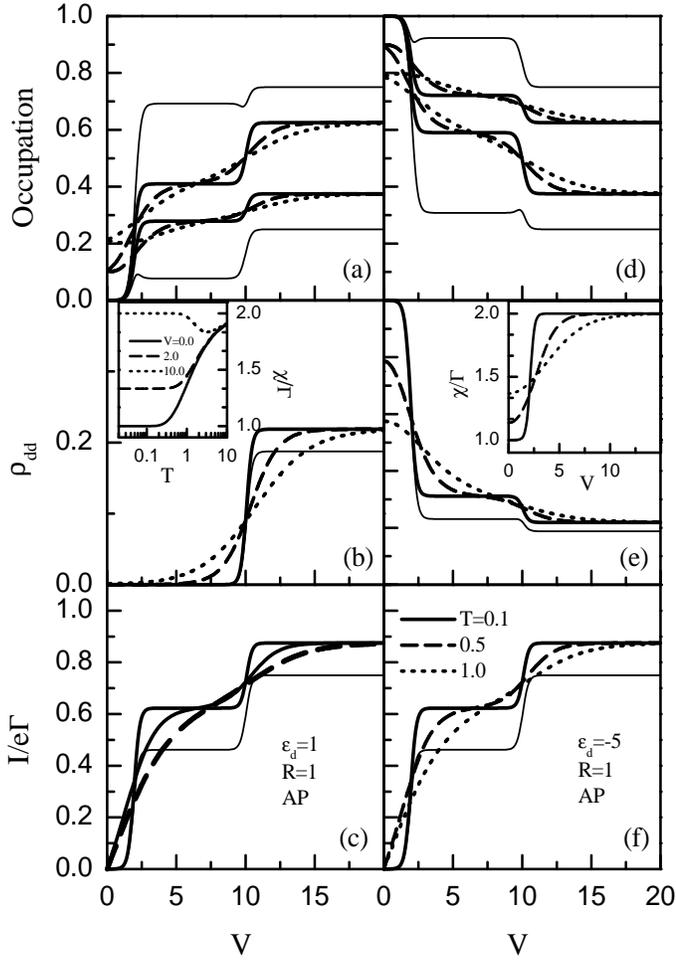}
\caption{Occupation numbers $n_{\uparrow}$, $n_{\downarrow}$ (a,d), $\rho_{dd}$ (b,e), 
and current (c,f) vs the bias voltage calculated for the AP configuration with the 
spin-flip transition $R=1$ and different temperatures $T=0.1$, $0.5$, and $1.0$. (a)-(c) 
are plotted for $\epsilon_d=1$, (d)-(f) for $\epsilon_d=-5$. For comparison, the 
respective results without the spin-flip transition are also plotted as thin lines. The 
insets in (b) and (e): the temperature and bias dependence of the spin relaxation rate. 
Other parameters are as in Fig.~2.}
\label{fig4}
\end{figure}

\begin{figure}[htb]
\includegraphics[height=5.in,width=3.in]{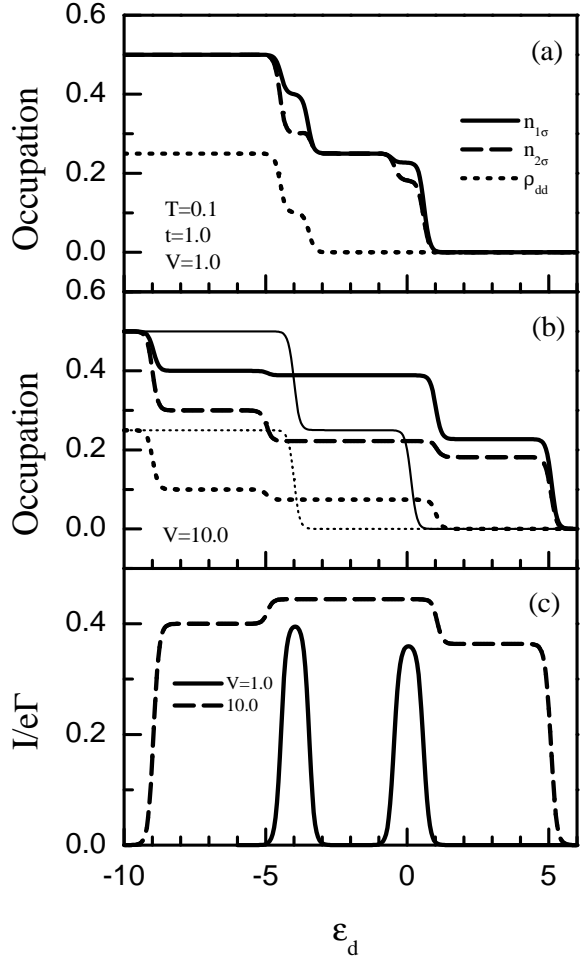}
\caption{Nonequilibrium occupation numbers $n_{1\sigma}$, $n_{2\sigma}$, $\rho_{dd}$ 
(a,b), and current (c) vs the bare level of the CQD. (a) is plotted at a small bias 
$V=1.0$ and (b) is at a large bias $V=10.0$. The equilibrium occupation numbers are also 
depicted by the thin lines in (b) and (e). Other parameters are: $U=4$, $T=0.1$, and 
$t=1.0$.}
\label{fig5}
\end{figure}

\begin{figure}[htb]
\includegraphics[height=5in,width=3.in]{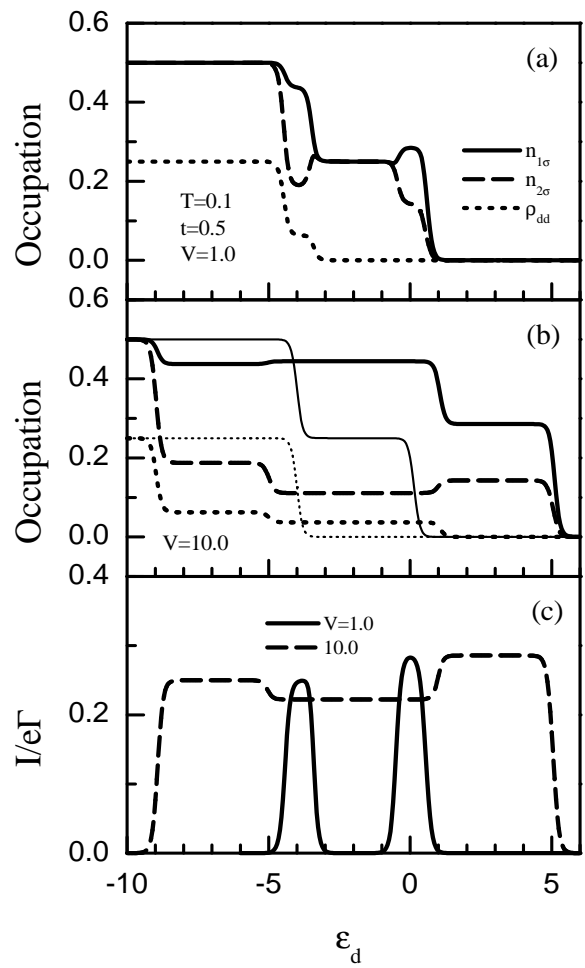}
\caption{The same as Fig.~5 except for $t=0.5$.}
\label{fig6}
\end{figure}

\begin{figure}[htb]
\includegraphics[height=5in,width=3.5in]{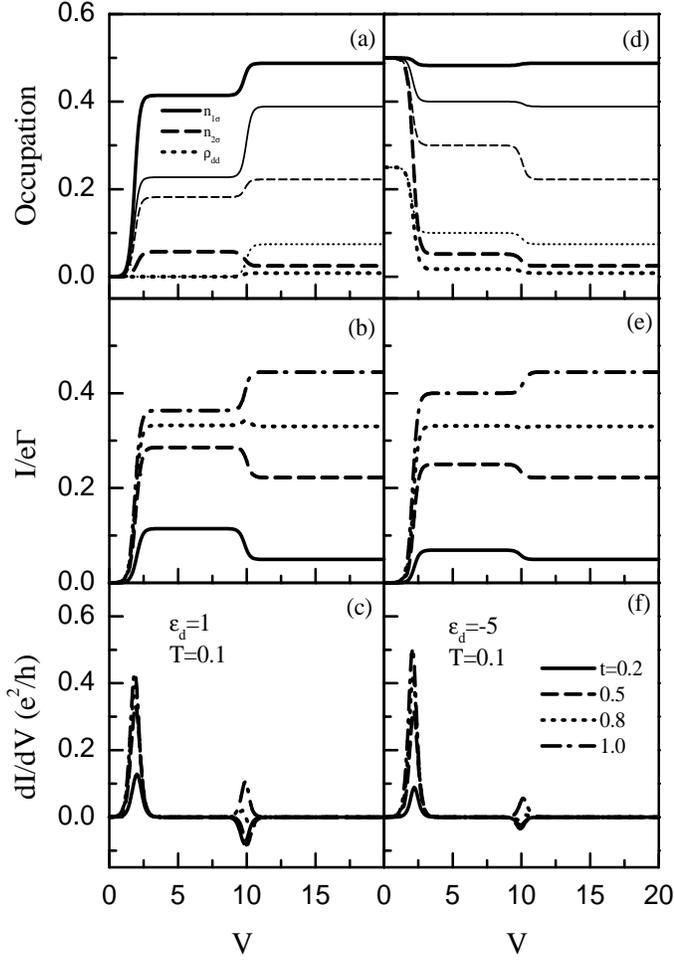}
\caption{Occupation numbers $n_{1\sigma}$, $n_{2\sigma}$, $\rho_{dd}$ (a,d), current 
(b,e), and the differential conductance (c,f) vs the bias voltage, calculated for 
different hopping $t$. (a)-(c) are plotted for $\epsilon_d=1$, (d)-(f) for 
$\epsilon_d=-5$. Other parameters are as in Fig.~5.}
\label{fig7}
\end{figure}

\begin{figure}[htb]
\includegraphics[height=2.5in,width=3.5in]{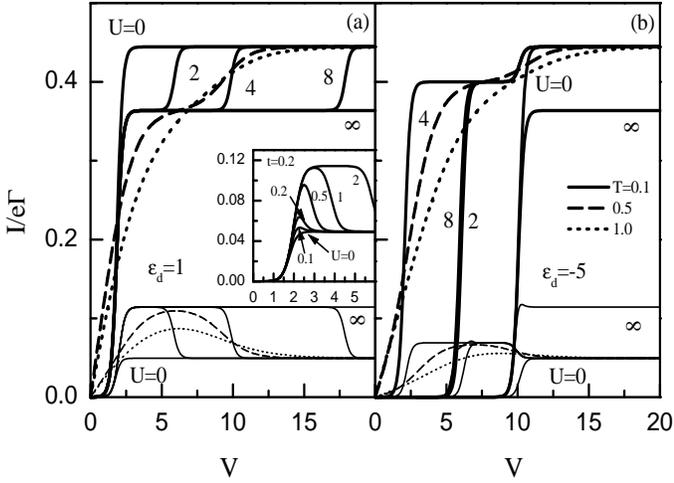}
\caption{The current-voltage characteristics, calculated for indicated values of the 
interdot Coulomb interaction $U$ and different hoppings $t=0.2$ (thin lines) and $1.0$ 
(thick lines). (a) is plotted for $\epsilon_d=1$, (b) is for $\epsilon_d=-5$. Inset in 
(a): $I$-$V$ curves for small $U$.}
\label{fig8}
\end{figure}

\begin{figure}[htb]
\includegraphics[height=3.in,width=3.5in]{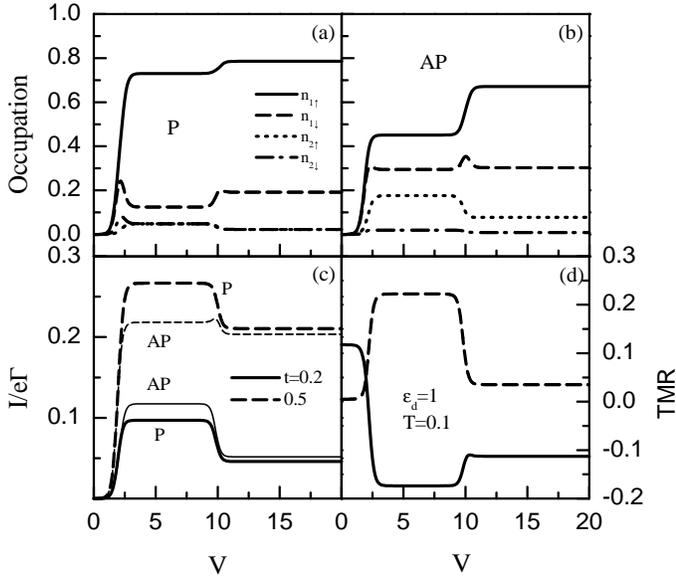}
\caption{Occupation numbers $n_{1\uparrow}$, $n_{1\downarrow}$, $n_{2\uparrow}$, and 
$n_{2\downarrow}$ in the P configuration (a) and the AP configuration (b) for $t=0.2$, 
current (c), and TMR (d) for $t=0.2$ and $0.5$ vs the bias voltage. Other parameters are 
$\epsilon_d=1$, $T=0.1$, and $p=0.5$.}
\label{fig9}
\end{figure}

\begin{figure}[htb]
\includegraphics[height=3in,width=3.5in]{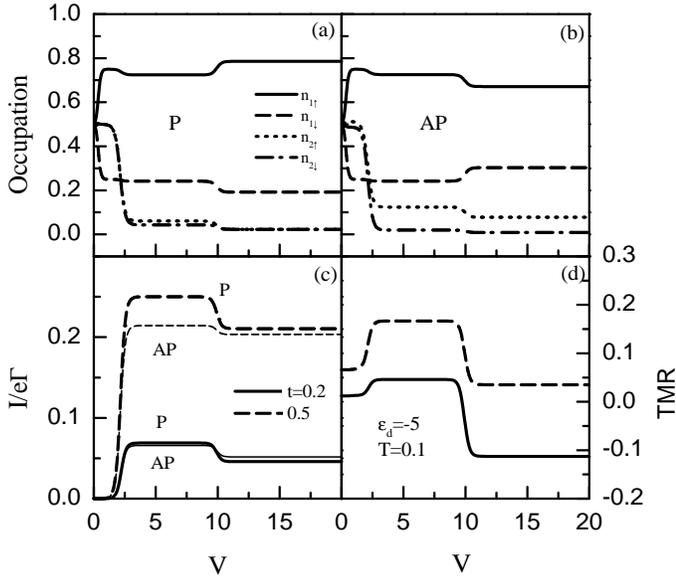}
\caption{The same as Fig.~9 but for the case of $\epsilon_d=-5$.}
\label{fig10}
\end{figure}

\end{document}